# Design-Based RCT Estimators and Central Limit Theorems for Baseline Subgroup and Related Analyses

**August 2023**

Peter Z. Schochet, Ph.D. (Corresponding Author)
Senior Fellow, Associate Director
Mathematica
P.O. Box 2393
Princeton, NJ 08543-2393
Phone: (609) 936-2783
pschochet@mathematica-mpr.com

## Abstract

There is a growing literature on design-based methods to estimate average treatment effects (ATEs) for randomized controlled trials (RCTs) for full sample analyses. This article extends these methods to estimate ATEs for discrete subgroups defined by pre-treatment variables, with an application to an RCT testing subgroup effects for a school voucher experiment in New York City. We consider ratio estimators for subgroup effects using regression methods, allowing for model covariates to improve precision, and prove a finite population central limit theorem. We discuss extensions to blocked and clustered RCT designs, and to other common estimators with random treatment-control sample sizes (or weights): post-stratification estimators, weighted estimators that adjust for data nonresponse, and estimators for Bernoulli trials. We also develop simple variance estimators that share features with robust estimators. Simulations show that the design-based subgroup estimators yield confidence interval coverage near nominal levels, even for small subgroups.

# 1. Introduction

There is a growing literature on design-based methods to estimate overall average treatment effects (ATEs) for randomized controlled trials (RCTs). These nonparametric methods use the building blocks of experimental designs to generate consistent, asymptotically normal ATE estimators with minimal assumptions. The underpinnings of these methods were introduced by Neyman [1] and later developed in seminal works by Rubin [2,3] and Holland [4] using a potential outcomes framework.

To date, the design-based literature has focused on ATE estimation for full sample analyses. In this article, we build on these methods to develop ATE estimators for discrete *subgroups* defined by pre-treatment (baseline) characteristics of study participants. Subgroup analyses for RCTs are common across fields as they can be used to assess treatment effect heterogeneity and inform decisions about how to best target and improve treatments [5,6]. Guidelines for the planning, analysis, and reporting of RCT subgroup analyses have been proposed in the literature to ensure statistical rigor, such as reducing the chances of finding spurious positive effects [5,7,8].

As a motivating example, consider the Evaluation of the New York City (NYC) School Choice Scholarships Program, an RCT where low-income public school students in grades K-4 could participate in a series of lotteries to receive a private school voucher for up to three years [9,10]. A subgroup analysis was pre-specified for the study to examine differences in voucher effects for African American and Latino students. The hypothesis was that African Americans might benefit more from the vouchers as they tended to live in poorer communities and attend lower-performing public schools.

Several key aspects of this subgroup analysis motivate the theory underlying this article. First, the study sample was not randomly sampled from a broader population. Rather, the sample included only a very small percentage of NYC families who applied for a scholarship. Thus, the study results cannot be generalized to a broader voucher program that would involve all children in NYC or elsewhere. This setting suggests a finite population framework for estimating ATEs where the sample and their potential outcomes are considered fixed, and study results are assumed to pertain to the study sample only. This is a common RCT setting across disciplines that often include volunteer samples of individuals and sites.

Second, the estimation strategy should allow for the inclusion of model baseline covariates to improve precision as power is often a concern for subgroup analyses due to small sample sizes. Third, the voucher study conducted randomization within strata, suggesting the need for a theory for blocked RCTs. Fourth, the study randomized families rather than students, suggesting a further need to consider a theory for clustered RCTs that are becoming increasingly prevalent across fields [11,12]. Finally, the study constructed weights to adjust for missing outcome data, a common strategy for RCT analyses that should be covered in the theory.

This article addresses these issues by developing design-based ATE ratio estimators for subgroup-related analyses using regression models that allow for baseline covariates. We focus on ratio estimators due to the randomness of subgroup sizes in the treatment and control groups. We prove a finite population central limit theorem (CLT) by building on the methods in Pashley [13] and Schochet et al. [14]. We also discuss extensions to blocked and clustered RCTs, and to other common estimators with random sample sizes (or weights): post-stratification estimators, weighted estimators that adjust for data nonresponse, and estimators for Bernoulli trials. We provide consistent variance estimators that are compared to commonly used robust standard errors. Our simulations show that the design-based subgroup ATE estimators yield confidence interval coverage near nominal levels, even for small subgroups. Finally, we demonstrate the methods using data from our motivating NYC voucher experiment.

The rest of this article proceeds as follows. Section 2 discusses the related literature. Section 3 provides the theoretical framework, ATE estimators, CLT for the non-clustered RCT, and extensions. Section 4 discusses blocked and clustered RCTs. Section 5 presents simulation results, and Section 6 presents empirical results using the voucher study. Section 7 concludes.

## 2. Related work

Our work builds on the growing literature on design-based methods to estimate ATEs for full sample analyses [14-23]. These methods also pertain to subgroup analyses conditional on subgroup sizes observed in the treatment and control groups [21], but not to unconditional analyses that average over subgroup allocations.

Our work draws most directly on two studies. First, we draw on methods in Schochet et al. [14] who provide finite-population CLTs for ratio estimators for blocked, clustered RCTs with general weights (using previous results in Scott and Wu [24], Li and Ding [23], and Pashley [13]). Our innovation is to adapt these methods by treating subgroup indicators as "weights" in the analysis. Second, we draw on results in Miratrix et al. [25] who consider design-based post-stratification estimators for overall effects, which share properties with baseline subgroup estimators. Miratrix et al. [25], however, do not consider asymptotic distributions, blocked or clustered RCT designs, the inclusion of other model covariates, or weights considered here.

Finally, there is a large statistical literature on design-based methods for analyzing survey data with complex sample designs, including for estimating subpopulation means or totals [26-28]. However, these works do not consider RCT settings for estimating treatment-control differences in subpopulation means.

In what follows, we focus on the non-clustered RCT design (without blocking) and extensions to related estimators. We then discuss blocked and clustered designs.

## 3. Design-based subgroup analysis for non-clustered RCTs

We assume an RCT of $n$ individuals, with $n^1 = np$ assigned to the treatment group and $n^0 = n(1-p)$ assigned to the control group, where $p$ is the treatment assignment rate ($0 < p < 1$). Let $Y_i(1)$ be the outcome of person $i$ if assigned to the treatment group and $Y_i(0)$ be the outcome in the control condition. These potential outcomes can be continuous, binary, or discrete. We assume a finite population model, where potential outcomes are assumed fixed for the study. Let $T_i$ equal 1 if person $i$ is randomly assigned to the treatment condition and 0 otherwise.

For the subgroup analysis, we assume each sample member is allocated to a discrete category within a subgroup class (such as an age group). The subgroup classes can be formed from continuous, categorical, or discrete variables measured at baseline, so are unaffected by the treatment. We consider estimation for each subgroup class in isolation. For a specific class, let $G_{ik}$ equal 1 for a member of subgroup (level) $k$ and 0 otherwise, for $k \in \{1, 2, \dots, K\}$ and $K \geq 1$. Let $n_k = \sum_{i=1}^{n} G_{ik}$ denote the number of persons in subgroup $k$, with $\sum_{k=1}^{K} n_k = n$. Finally, let $\pi_k = \bar{G}_k = n_k/n$ be the subgroup population share, with $\sum_{k=1}^{K} \pi_k = 1$.

We assume two conditions. The first is the stable unit treatment value assumption (SUTVA) [29]:

*(C1): SUTVA*: Let $Y_i(\mathbf{T})$ denote the potential outcome given the random vector of all treatment assignments, $\mathbf{T}$. Then, if $T_i = T_i'$ for person $i$, we have that $Y_i(\mathbf{T}) = Y_i(\mathbf{T}')$.

SUTVA allows us to express $Y_i(\mathbf{T})$ as $Y_i(T_i)$, so that a person's potential outcomes depend only on the person's treatment assignment and not on those of other persons in the sample. This condition is assumed to hold within and across subgroups. SUTVA also assumes a particular treatment unit cannot receive different forms of the treatment.

Under SUTVA, the ATE parameter for subgroup $k$ under the finite population model is,

$$\tau_k = \frac{\sum_{i=1}^n G_{ik}(Y_i(1) - Y_i(0))}{n_k} = \bar{Y}_k(1) - \bar{Y}_k(0), \tag{1}$$

which is the mean treatment effect for members of subgroup $k$.

Our second condition is complete randomization [21], where extensions to Bernoulli trials are discussed in Remark (8) in Section 3.2:

*(C2): Complete randomization*: For fixed $n^1$, if $\mathbf{t} = (t_1, \ldots, t_n)$ is any vector of randomization realizations such that $\sum_{i=1}^n t_i = n^1$, then $Prob(\mathbf{T} = \mathbf{t}) = \binom{n}{n^1}^{-1}$.

This condition implies that potential outcomes are independent of treatment status, $Y_i(1), Y_i(0) \perp\!\!\!\perp T_i$, which also holds for any baseline subgroup (e.g., males or females).

### 3.1. ATE estimators

Under the potential outcomes framework and SUTVA, the data generating process for the observed outcome measure, $y_i$, is a result of the random assignment process:

$$y_i = T_i Y_i(1) + (1 - T_i) Y_i(0). \tag{2}$$

This relation states that we can observe $y_i = Y_i(1)$ for those in the treatment group and $y_i = Y_i(0)$ for those in the control group, but not both.

Rearranging (2) generates the following nominal full sample regression model:

$$y_i = \alpha + \tau \tilde{T}_i + u_i, \tag{3}$$

where $\tilde{T}_i = (T_i - p)$ is the centered treatment indicator; $\tau = \bar{Y}(1) - \bar{Y}(0)$ is the full sample ATE estimand; $\bar{Y}(t) = \frac{1}{n}\sum_{i=1}^n Y_i(t)$ is the mean potential outcome for $t \in \{1,0\}$; $\alpha = p\bar{Y}(1) + (1-p)\bar{Y}(0)$ is the intercept (expected outcome); and the "error" term, $u_i$, is,

$$u_i = T_i(Y_i(1) - \bar{Y}(1)) + (1 - T_i)(Y_i(0) - \bar{Y}(0)).$$

We center the treatment indicator in (3) to facilitate the theory without changing the estimator.

In contrast to usual formulations of the regression model, the residual, $u_i$, is random solely due to $T_i$ [16,17,20]. This framework allows individual-level treatment effects, $\tau_i = Y_i(1) - Y_i(0)$, to vary across the sample, and is nonparametric because it makes no assumptions about the potential outcome distributions. The model does not satisfy key assumptions of the usual regression model: over the randomization distribution ($R$), $u_i$ is heteroscedastic, and $E_R(u_i)$, $Cov_R(u_i, u_{i'})$, and $E_R(\tilde{T}_i u_i)$ are nonzero if $\tau_i$ varies across the sample.

The model in (3) also applies to each subgroup due to randomization. Thus, if we combine each subgroup model using the $G_{ik}$ indicators, we obtain the following pooled model:

$$y_i = \sum_{k=1}^K \tau_k G_{ik} \tilde{T}_i + \sum_{k=1}^K \alpha_k G_{ik} + \epsilon_i, \tag{4}$$

where $\tau_k = \bar{Y}_k(1) - \bar{Y}_k(0)$ is the subgroup ATE estimand; $\alpha_k = p\bar{Y}_k(1) + (1-p)\bar{Y}_k(0)$ is the subgroup intercept; and $\epsilon_i = \sum_{k=1}^K G_{ik} u_{ik}$ is the error term, with $u_{ik} = T_i(Y_i(1) - \bar{Y}_k(1)) + (1 - T_i)(Y_i(0) - \bar{Y}_k(0))$. We include model terms for all subgroups and exclude the grand intercept.

Consider the ordinary least squares (OLS), differences-in-mean estimator for $\tau_k$ from (4) using data on the full sample:

$$\hat{\tau}_k = \bar{y}_k^1 - \bar{y}_k^0 = \frac{1}{n_k^1}\sum_{i=1}^n G_{ik} T_i Y_i(1) - \frac{1}{n_k^0}\sum_{i=1}^n G_{ik}(1 - T_i)Y_i(0), \tag{5}$$

where $n_k^1 = \sum_{i=1}^n T_i G_{ik}$ and $n_k^0 = \sum_{i=1}^n (1-T_i) G_{ik}$ are subgroup sizes in the treatment and control groups, with sample shares, $\pi_k^1 = n_k^1/n^1$ and $\pi_k^0 = n_k^0/n^0$. We see that $\hat{\tau}_k$ is a ratio estimator because $n_k^1$ and $n_k^0$ are random variables (with hypergeometric distributions).

Our theory is conditional on randomizations that yield $n_k^1 > 0$ and $n_k^0 > 0$ so that $\hat{\tau}_k$ and its variance can be defined [25]. However, these restrictions (not denoted for simplicity), have little effect on our results as they will hold with probability near 1 for typical subgroup analyses. For instance, even for a very small subgroup with $n_k = 12$, $n = 40$, and $p = .5$, the restrictions will hold with probability 0.9999. Thus, to reduce notation, we ignore the slight truncation effects on subgroup treatment assignment probabilities (that converge to zero as $n$ and $n_k$ increase), and we assume that $n_k^1$ and $n_k^0$ have unrestricted expectations, $n_k p$ and $n_k(1-p)$.

The finite population CLT in Theorem 4 in Li and Ding [23] applies to $\hat{\tau}_k$ conditional on observed subgroup allocations. However, a complete design-based theory requires an unconditional CLT. We provide this CLT in Section 3.2. for a more general covariate-adjusted estimator from a working model that includes in (4) a $1xV$ vector of fixed, centered baseline covariates (other than the subgroup indicators), $\tilde{\mathbf{x}}_{ik} = (\mathbf{x}_i - \bar{\mathbf{x}}_k)$, with parameter vector, $\boldsymbol{\beta}$:

$$y_i = \sum_{k=1}^K \tau_k G_{ik} \tilde{T}_i + \sum_{k=1}^K \alpha_k G_{ik} + \tilde{\mathbf{x}}_{ik} \boldsymbol{\beta} + e_i, \qquad (6)$$

where $\bar{\mathbf{x}}_k = \frac{1}{n_k} \sum_{i=1}^n G_{ik} \mathbf{x}_i$ are covariate means and $e_i$ is the error term. While the covariates do not enter the true RCT model in (4) and the ATE estimands do not change, they will increase precision to the extent they are correlated with the potential outcomes. We do not need to assume that the true conditional distribution of $y_i$ given $\mathbf{x}_i$ is linear in $\mathbf{x}_i$. We define $\boldsymbol{\beta}$ in Section 3.2.

We focus on the pooled covariate model in (6) because it is commonly used in practice. In Remark 7 in Section 3.2, we discuss extensions to models that interact $\tilde{T}_i$ with $\tilde{\mathbf{x}}_{ik}$ and $G_{ik}$.

Using OLS to estimate the working model in (6) yields the following covariate-adjusted estimator for $\tau_k$ that is produced by standard OLS statistical packages:

$$\hat{\tau}_k^x = (\bar{y}_k^1 - \bar{y}_k^0) - (\bar{\mathbf{x}}_k^1 - \bar{\mathbf{x}}_k^0)\hat{\boldsymbol{\beta}}, \tag{7}$$

where $\bar{\mathbf{x}}_k^1$ and $\bar{\mathbf{x}}_k^0$ are subgroup covariate means for treatments and controls, and $\hat{\boldsymbol{\beta}}$ is the OLS estimator for $\boldsymbol{\beta}$ (see [23] for a parallel result for full sample analyses).

### 3.2. CLT result and extensions

To consider the asymptotic properties of $\hat{\tau}_k^x$ (which also apply to $\hat{\tau}_k$), we consider a hypothetical increasing sequence of finite populations where $n \to \infty$. Parameters should be subscripted by $n$, but we omit this notation for simplicity. We assume that $n^1/n \to p^*$ as $n \to \infty$, so the numbers of treatments and controls both increase with $n$. In addition, we assume that $n_k/n \to \pi_k^*$ for all $k$, where $\pi_k^* > 0$ and $\sum_{k=1}^{K} \pi_k^* = 1$. This implies that each subgroup also grows with $n$.

Our CLT builds on Schochet et al. [14] who provide CLTs for RCT ratio estimators with general weights for clustered, blocked designs. We adapt these methods to our setting by treating the subgroup indicators, $G_{ik}$, as "weights" when computing the subgroup sample means.

Before presenting our CLT, we need to define several terms. First, for $t \in \{1,0\}$, let $\varepsilon_{ik}(t) = (Y_i(t) - \bar{Y}_k(t) - \tilde{\mathbf{x}}_{ik}\boldsymbol{\beta})$ denote model residuals for subgroup $k$, where $R_{ik}(t) = \frac{G_{ik}}{\pi_k}\varepsilon_{ik}(t)$ are scaled residuals using the normalized weights, $w_i/\bar{w} = G_{ik}/\pi_k$, that sum to $n$. Second, let $S_{R_k}^2(t) = \frac{1}{n-1}\sum_{i=1}^{n} R_{ik}^2(t)$ denote the variance of $R_{ik}(t)$, and let $S_{R_k}^2(1,0) = \frac{1}{n-1}\sum_{i=1}^{n} R_{ik}(1)R_{ik}(0)$ denote the treatment-control covariance. Third, we define $\bar{D}_k$ as the mean treatment-control difference in the $R_{ik}(t)$ residuals, with associated variance:

$$\text{Var}(\bar{D}_k) = \frac{S_{R_k}^2(1)}{n^1} + \frac{S_{R_k}^2(0)}{n^0} - \frac{S^2(\tau_k)}{n}, \tag{8}$$

where $S^2(\tau_k) = \frac{1}{n-1}\sum_{i=1}^n (R_{ik}(1) - R_{ik}(0))^2$ is the heterogeneity of treatment effects. Fourth,

we define the variance of $G_{ik}$ as $S^2(G_k) = \frac{1}{n-1}\sum_{i=1}^n \frac{1}{\pi_k^2}(G_{ik} - \pi_k)^2 = \frac{n}{(n-1)}\frac{(1-\pi_k)}{\pi_k}$. Fifth, we require the variances of each covariate, $S^2_{x_k,v} = \frac{1}{n-1}\sum_{i=1}^n \frac{G_{ik}}{\pi_k^2}([\tilde{\mathbf{x}}_{ik}]_v)^2$ for $v \in \{1, \ldots, V\}$, and the full variance-covariance matrix for the covariates, $\mathbf{S}^2_{\mathbf{x},k} = \frac{1}{n}\sum_{i=1}^n \frac{G_{ik}}{\pi_k}\tilde{\mathbf{x}}'_{ik}\tilde{\mathbf{x}}_{ik}$. Finally, we need two outcome-covariate variance-covariance matrices: $\mathbf{S}^2_{\mathbf{x},Y,k}(t) = \frac{1}{n}\sum_{i=1}^n \frac{G_{ik}}{\pi_k}\tilde{\mathbf{x}}'_{ik}Y_i(t)$ and $\mathbf{S}^2_{\mathbf{x}Y,k}(t) =$

$\frac{1}{n}\sum_{i=1}^n \frac{1}{\pi_k}(G_{ik}\tilde{\mathbf{x}}'_{ik}Y_i(t) - \bar{\boldsymbol{\theta}}_k)^2$, where $\bar{\boldsymbol{\theta}}_k = \frac{1}{n}\sum_{i=1}^n G_{ik}\tilde{\mathbf{x}}'_{ik}Y_i(t)$ is the mean covariance.

We now present our CLT theorem, proved in Supplementary Materials A.

**Theorem 1.** Assume (*C1*), (*C2*), and the following conditions for $t \in \{1,0\}$, $k \in \{1, \ldots, K\}$, and $K \geq 1$:

(*C3*) Letting $g_k(t) = \max_{1 \leq i \leq n}\{R_{ik}^2(t)\}$, as $n \to \infty$,

$$\frac{1}{(n^t)^2}\frac{g_k(t)}{\text{Var}(\bar{D}_k)} \to 0.$$

(*C4*) $f^1 = n^1/n$ and $f^0 = n^0/n$ have limiting values, $p^*$ and $(1-p^*)$, for $0 < p^* < 1$.

(*C5*) The subgroup shares, $n_k/n$, converge to $\pi_k^*$ for $0 < \pi_k^* < 1$ and $\sum_{k=1}^K \pi_k^* = 1$.

(*C6*) As $n \to \infty$,

$$(1-f^t)\frac{S^2(G_k)}{n^t} \to 0.$$

(*C7*) Letting $h_v(t) = \max_{1 \leq i \leq n}\left\{\frac{G_{ik}}{\pi_k}([\tilde{\mathbf{x}}_{ik}]_v)\right\}^2$ for all $v \in \{1, \ldots, V\}$, as $n \to \infty$,

$$\frac{1}{\min(n^1, n^0)}\frac{h_v(t)}{S^2_{x,v}} \to 0.$$

(*C8*) $S^2_{R_k}(t)$, $S^2_{R_k}(1,0)$, $S^2_{x_k,v}$, $\mathbf{S}^2_{\mathbf{x},k}$, $\mathbf{S}^2_{\mathbf{x},Y,k}(t)$, and $\mathbf{S}^2_{\mathbf{x}Y,k}(t)$ have finite limiting values.

Then, as $n \to \infty$, $\hat{\tau}_k^x$ is a consistent estimator for $\tau_k$, and

$$\frac{\hat{\tau}_k^x - (\overline{Y}_k(1) - \overline{Y}_k(0))}{\sqrt{\text{Var}(\overline{D}_k)}} \xrightarrow{d} N(0,1),$$

where $\text{Var}(\overline{D}_k)$ is defined as in (8).

*Remark 1.* The $\text{Var}(\overline{D}_k)$ expression in (8) is difficult to interpret because $S_{R_k}^2(t)$ and $S^2(\tau_k)$ are scaled by $\pi_k^2(n-1)$ to facilitate the theory. To address this, we apply the following relations in (8): $n^1 \pi_k^2(n-1) = n_k p(n_k - \pi_k)$ and $n^0 \pi_k^2(n-1) = n_k(1-p)(n_k - \pi_k)$, which yields,

$$\text{Var}(\overline{D}_k) = \phi_k \left[ \frac{\Omega_{R_k}^2(1)}{n_k p} + \frac{\Omega_{R_k}^2(0)}{n_k(1-p)} - \frac{\Omega^2(\tau_k)}{n_k} \right], \tag{9}$$

where $\Omega_{R_k}^2(t) = \frac{1}{n_k - 1} \sum_{i=1}^n G_{ik} \varepsilon_{ik}^2(t)$; $\Omega^2(\tau_k) = \frac{1}{n_k - 1} \sum_{i=1}^n G_{ik}(\varepsilon_{ik}(1) - \varepsilon_{ik}(0))^2$; and $\phi_k = (n_k - 1)/(n_k - \pi_k) \leq 1$ is a correction term that reflects the single treatment indicator "shared" by each subgroup (and can be ignored as it converges to 1). The $\Omega_{R_k}^2(t)$ and $\Omega^2(\tau_k)$ terms are population variances for those in subgroup $k$, and $n_k p$ and $n_k(1-p)$ are expected subgroup sizes in the two research groups. This variance expression is more intuitive as it parallels the full sample asymptotic results in Li and Ding [23], the key difference being that (9) is based on expected subgroup sizes rather than actual ones. Note that for $\phi_k = 1$, (9) is the same as for an RCT that stratifies on subgroup $k$ to select fixed subgroup sample sizes, $n_k p$ and $n_k(1-p)$.

*Remark 2.* The first two terms in (9) pertain to separate variances for the two research groups because we allow for heterogeneous treatment effects. The third term pertains to the treatment-control covariance, $\Omega_{R_k}^2(1,0)$, expressed in terms of the heterogeneity of treatment effects, $\Omega^2(\tau_k)$, which cannot be identified from the data but can be bounded [30].

*Remark 3.* (C3) and (C7) are Lindeberg-type conditions from Li and Ding [23] that control the tails of the potential outcome and covariate distributions. (C6) yields a weak law of large numbers for the observed subgroup shares so that $\pi_k^t / \pi_k \xrightarrow{p} 1$ (using Theorem B in Scott and Wu [24]). While (*C6*) is implied by *(C4)* and (*C5*), it facilitates the addition of other weights (see Remark 9). (*C8*) specifies limiting values of the variances and variance-covariance matrices.

*Remark 4.* Theorem 1 is proved in two stages by expressing the ATE estimator in (7) as,

$$\hat{\tau}_k^{\text{x}} = \hat{\tau}_k^{\text{x}\beta} - (\bar{\mathbf{x}}_k^1 - \bar{\mathbf{x}}_k^0)(\hat{\boldsymbol{\beta}} - \boldsymbol{\beta}), \tag{10}$$

where $\hat{\tau}_k^{\text{x}\beta} = (\bar{y}_k^1 - \bar{y}_k^0) - (\bar{\mathbf{x}}_k^1 - \bar{\mathbf{x}}_k^0)\boldsymbol{\beta}$, and $\boldsymbol{\beta} = \left(\sum_{k=1}^K \pi_k \mathbf{S}_{\mathbf{x},k}^2\right)^{-1}[\sum_{k=1}^K p\pi_k \mathbf{S}_{\mathbf{x},Y,k}^2(1) + \sum_{k=1}^K (1-p)\pi_k \mathbf{S}_{\mathbf{x},Y,k}^2(0)]$ is assumed known. This $\boldsymbol{\beta}$ parameter is the (hypothetical) population OLS coefficient that would result from a regression of $[pY_i(1) + (1-p)Y_i(0)]$ on the covariates. In the first stage, we obtain a CLT for $\hat{\tau}_k^{\text{x}\beta}$. In the second stage, we prove that $\hat{\tau}_k^{\text{x}}$ has the same asymptotic distribution as $\hat{\tau}_k^{\text{x}\beta}$ by showing that $(\bar{\mathbf{x}}_k^1 - \bar{\mathbf{x}}_k^0)(\hat{\boldsymbol{\beta}} - \boldsymbol{\beta}) = o_p(n^{-1/2})$, which holds under our conditions because $\bar{\mathbf{x}}_k^1$ and $\bar{\mathbf{x}}_k^0$ are both asymptotically normal and $\hat{\boldsymbol{\beta}} - \boldsymbol{\beta} \xrightarrow{p} \mathbf{0}$.

*Remark 5.* As proved in Corollary 1 to Theorem 1 in Supplement A, the subgroup estimators, $(\hat{\tau}_1^{\text{x}}, \ldots, \hat{\tau}_K^{\text{x}})$, are asymptotically independent with a joint multivariate normal distribution. Thus, standard F-tests (or chi-square tests) can be used to test the null of equal subgroup effects.

*Remark 6.* Theorem 1 can be extended to a model that replaces $\tilde{\mathbf{x}}_{ik}\boldsymbol{\beta}$ in (6) with the interaction terms, $\sum_{k=1}^K G_{ik}(1-T_i)\tilde{\mathbf{x}}_{ik}\boldsymbol{\beta}_k^0$ and $\sum_{k=1}^K G_{ik}T_i\tilde{\mathbf{x}}_{ik}\boldsymbol{\beta}_k^1$, which allows covariate effects to differ by subgroup and treatment status. The ATE estimator for this model is, $\hat{\tau}_k^{\text{x}GT} = [\bar{y}_k^1 - (\bar{\mathbf{x}}_k^1 - \bar{\mathbf{x}}_k)\hat{\boldsymbol{\beta}}_k^1] - [\bar{y}_k^0 - (\bar{\mathbf{x}}_k^0 - \bar{\mathbf{x}}_k)\hat{\boldsymbol{\beta}}_k^0]$. Theorem 1 can then be applied by redefining the residuals as $R_{ik}(t) = \frac{G_{ik}}{\pi_k}(Y_i(t) - \bar{Y}_k(t) - (\mathbf{x}_i - \bar{\mathbf{x}}_k)\boldsymbol{\beta}_k^t)$, where $\boldsymbol{\beta}_k^t = (\mathbf{S}_{\mathbf{x},k}^2)^{-1}\mathbf{S}_{\mathbf{x},Y,k}^2(t)$. The proof follows using the same arguments as in Remark 3 by replacing (10) with $\hat{\tau}_k^{\text{x}GT} = \hat{\tau}_k^{\text{x}GT\beta} - \sum_{t=0}^1(\bar{\mathbf{x}}_k^t - \bar{\mathbf{x}}_k)(\hat{\boldsymbol{\beta}}_k^t - \boldsymbol{\beta}_k^t)$, and noting that under our regularity conditions, $\hat{\boldsymbol{\beta}}_k^t - \boldsymbol{\beta}_k^t \xrightarrow{p} \mathbf{0}$ and $(\bar{\mathbf{x}}_k^t - \bar{\mathbf{x}}_k)(\hat{\boldsymbol{\beta}}_k^t - \boldsymbol{\beta}_k^t) = o_p(n^{-1/2})$ for $t \in \{1,0\}$, so that $\sum_{t=0}^1(\bar{\mathbf{x}}_k^t - \bar{\mathbf{x}}_k)(\hat{\boldsymbol{\beta}}_k^t - \boldsymbol{\beta}_k^t) = o_p(n^{-1/2})$.

*Remark 7.* Under (C1)-(C6), Theorem 1 also applies to $\hat{\tau}_k$ for the model without covariates by setting $\boldsymbol{\beta} = \mathbf{0}$. In Supplement B.1, we prove that $\hat{\tau}_k$ is unbiased using the approach in Miratrix et al. [25] and Schochet [18], where we condition on $n_k^1 > 0$ and $n_k^0 > 0$ and then average over possible subgroup allocations ($A$) to the two research groups to show that $E_R(\hat{\tau}_k) = E_A E_R(\hat{\tau}_k | n_k^1, n_k^0) = \tau_k$. Similarly, using the law of total variance, Supplement B.1 shows that,

$$\text{Var}_R(\bar{D}_k) = E_A\left(\frac{1}{n_k^1}\right)\Omega_{R_k}^2(1) + E_A\left(\frac{1}{n_k^0}\right)\Omega_{R_k}^2(0) - \frac{\Omega^2(\tau_k)}{n_k}, \quad (11)$$

where $\Omega_{R_k}^2$ and $\Omega^2$ are defined in (9) with $\boldsymbol{\beta} = \mathbf{0}$. Note that $\lim_{n\to\infty} E_A\left(\frac{1}{n_k^1}\right) = \frac{1}{E_A(n_k^1)} = \frac{1}{n_k p}$ and $\lim_{n\to\infty} E_A\left(\frac{1}{n_k^0}\right) = \frac{1}{n_k(1-p)}$ as used in (9). In finite samples, however, $E_A\left(\frac{1}{n_k^1}\right) \geq \frac{1}{n_k p}$ and $E_A\left(\frac{1}{n_k^0}\right) \geq \frac{1}{n_k(1-p)}$ by Jensen's inequality. Our simulations include both sets of sample sizes (Section 5).

*Remark 8.* The results in Remark 7 extend to Bernoulli trials where each sample member is independently randomized to the treatment group with probability $p$, leading to random treatment-control sizes. This design pertains, for example, to an RCT with rolling study intake. The only change from Remark 7 is that $E_A$ is now taken over a (truncated) binomial rather than hypergeometric distribution. To see this, consider the full sample analysis. Then, Bernoulli sampling has the same properties as a two-stage design that first randomly selects $n^1$ from a binomial distribution (assuming as above that $0 < n^1 < n$ with certainty), and then selects a simple random sample of size $n^1$ to the treatment group [31]. Thus, this setting is parallel to the one in Remark 7 that calculates sample moments by first conditioning on subgroup sizes.

We can also adapt Theorem 1 to Bernoulli trials by using expected rather than actual sizes in the theorem conditions. To see this, consider again the full sample analysis, omitting the $k = 1$ subscript for simplicity. Let $p^1$ be the observed treatment share, and express the mean outcomes as, $\bar{y}^1 = \bar{y}_{BT}^1 g^1$ and $\bar{y}^0 = \bar{y}_{BT}^0 g^0$, where $\bar{y}_{BT}^1 = \frac{1}{np}\sum_{i=1}^n T_i y_i$ and $\bar{y}_{BT}^0 = \frac{1}{n(1-p)}\sum_{i=1}^n (1-T_i)y_i$ are divided by expected sizes, with $g^1 = \frac{p}{p^1}$ and $g^0 = \frac{(1-p)}{(1-p^1)}$. Note that $g^t \xrightarrow{p} 1$ for $t \in \{1, 0\}$, so $\bar{y}^t$ and $\bar{y}_{BT}^t$ have the same asymptotic distributions. Then, under our revised conditions, Theorem 4 in [23] provides a CLT for $(\bar{y}_{BT}^1 - \bar{y}_{BT}^0)$, and the same proof as in Supplement A.2 for Theorem 1 extends this CLT to $(\bar{y}^1 - \bar{y}^0)$. A similar approach yields a CLT for the covariate-adjusted ATE estimator using the variance in (9) (by applying $\bar{\mathbf{x}}^t = \bar{\mathbf{x}}_{BT}^t g^t$ and noting that the full sample properties of $\hat{\boldsymbol{\beta}}$ do not change), and similarly for the subgroup analysis.

*Remark 9.* Theorem 1 also applies to a "subgroup" analysis that adjusts for missing outcome data using respondent outcomes only with weights, $w_i^r$. Let $r_i = R_i(T_i) = 1$ for respondents and 0 for nonrespondents, where $R_i(T_i)$ are potential responses. If baseline covariates are available for all, a common approach is to set $w_i^r = 1/e(\mathbf{x}_i)$, where $e(\mathbf{x}_i) = Pr(r_i = 1 | \mathbf{x}_i)$ is the propensity score [32]. We assume $e(\mathbf{x}_i)$ converges to $e^*(\mathbf{x}_i)$ as $n \to \infty$, where $0 < e^*(\mathbf{x}_i) \le 1$ for all $\mathbf{x}_i$ in its finite population support (so (C6) holds). We also assume for each subgroup that (i) data are missing at random [33], where $Y_i(1), Y_i(0) \perp\!\!\!\perp r_i | T_i = t, G_{ik} = 1, \mathbf{x}_i$; and (ii) data response and treatment status are independent conditional on the covariates: $r_i \perp\!\!\!\perp T_i | G_{ik} = 1, \mathbf{x}_i$. The weighting classes can then be treated as "pre-treatment" subgroups because potential responses are fixed, observed, and unaffected by the treatment.

Accordingly, we can apply Theorem 1, assuming known weights, where the respondent sample and weighted least squares (WLS) are used to obtain the ATE estimator, $\hat{\tau}_k^{rx}$, using (6). For the proof, we replace the weights, $w_i = G_{ik}$, in the theorem with $w_i = G_{ik} r_i w_i^r$ to define the variables and regularity conditions. The resulting variance for the CLT has the same form as (9) but is based on expected subgroup respondent sizes (see Supplement B.2). Developing a finite population CLT that allows for estimated nonresponse weights is a topic for future research.

### 3.3. Variance estimation

To obtain consistent variance estimators for (9) (and model variants), we can either use expected subgroup sizes, $n_k p$ and $n_k(1-p)$, or actual ones, $n_k^1$ and $n_k^0$ (as for the conditional analysis). Our simulations find very similar results using either approach (Section 5). This occurs because the difference between the hypergeometric random variable, $\pi_k^t$, and its expected value, $\pi_k$, decreases exponentially with $n^t$ [34,35]. For instance, Figure 1A shows that for modest $n^t$, there is a high probability that $|\pi_k^t - \pi_k|/\pi_k \leq c$ for small $c$ (defined as 10 or 20 percent).

Further, Figure 1B shows that for small $c$, the ratios of standard errors (SEs) using actual to expected subgroup sizes in (9) are close to 1, leading to similar confidence interval coverage. For example, for $n = 100$, $p = .5$, and $\pi_k = .5$, the ratios range only from 1 to 1.026 as $c$ ranges from 0 to .2 (assuming $\Omega_{R_k}^2(1) = \varphi \Omega_{R_k}^2(0)$ and $\Omega^2(\tau_k) = \vartheta \Omega_{R_k}^2(0)$ with plausible values, $\varphi = 1.1$ and $\vartheta = .05$). In expectation, the SE ratios are *greater* than 1 for all values of $\varphi$ and $\vartheta$ (see Remark 7 above), but the differences are small for typical subgroup sizes used in practice.

To further examine the SE ratios in Figure 1B, suppose first that $\varphi = 1$. Then, all ratios are at least 1 when $p = .5$. However, if $p < .5$, the ratios are greater than 1 if $\delta^1 = n_k^1 - n_k p < 0$ or $\delta^1 > n_k(1 - 2p)$, but are less than 1 otherwise, and vice versa for $p > .5$. As a function of $\delta^1$, the ratios are convex and symmetric around their minimum value at $\delta^1 = .5n_k(1 - 2p)$ when $n_k^1 = n_k^0$. This symmetry is lost when $\varphi \neq 1$, but the same overall patterns apply (Figure 1B).

Using expected sizes, a consistent (upper bound) plug-in variance estimator for (9) based on estimated subgroup regression residuals is as follows:

$$\text{Vâr}(\bar{D}_k) = \frac{s_{R_k}^2(1)}{n_k p} + \frac{s_{R_k}^2(0)}{n_k(1-p)}, \tag{12}$$

where

$$s_{R_k}^2(1) = \frac{1}{(n_k^1 - Vp\pi_k^1 - 1)} \sum_{i=1}^n T_i G_{ik} \left(y_i - \hat{\alpha}_k^x - (1-p)\hat{\tau}_k^x - \tilde{\mathbf{x}}_{ik}\hat{\boldsymbol{\beta}}\right)^2$$

and

$$s_{R_k}^2(0) = \frac{1}{(n_k^0 - V(1-p)\pi_k^0 - 1)} \sum_{i=1}^n (1-T_i) G_{ik} \left(y_i - \hat{\alpha}_k^x + p\hat{\tau}_k^x - \tilde{\mathbf{x}}_{ik}\hat{\boldsymbol{\beta}}\right)^2.$$

Here, we set $\phi_k = 1$, which can be relaxed by subtracting $\pi_k^t$ in the denominators of $s_{R_k}^2(t)$ rather than 1. In (12), the losses in degrees of freedom (*df*) due to the $V$ covariates are split proportionately across the $K$ subgroups and two research conditions. Note that the same estimator results using a non-centered model in (6) that replaces the $G_{ik}\tilde{T}_i$ and $\tilde{\mathbf{x}}_i$ terms with $G_{ik}T_i$ and $\mathbf{x}_i$. Hypothesis testing can be conducted using t-tests with $df = (n_k - V\pi_k - 2)$ or z-tests. Using actual sizes, we can instead use $n_k^1$ and $n_k^0$ in (12) rather than $n_k p$ and $n_k(1-p)$.

As shown in Supplement C, (12) is asymptotically equivalent to the robust Huber-White (HW) variance estimator [36,37], as has been shown for full sample estimators [16,20,38]. In finite samples, however, the design-based variances will typically be larger for the model without covariates due to larger *df* corrections We compare the two estimators in our simulations, along with other standard error variants.

## 4. Blocked and clustered designs

The above CLT results extend directly to blocked RCTs where randomization is performed separately within strata (e.g., sites, demographic groups, or time cohorts), and to clustered RCTs where groups (e.g., schools, hospitals, or communities) are randomized rather than individuals.

*4.1. Blocked RCTs*

In blocked designs, the sample is first divided into subpopulations, and a mini-experiment is conducted in each one. Note that we do not consider blocks formed by subgroups slated for ATE estimation as the theory for the full sample analysis applies in this case (as $n_k^1$ and $n_k^0$ are fixed).

For the blocked design, we use similar notation as above with the addition of the subscript $b = (1, 2, ..., h)$ to indicate blocks. For instance, $T_{ib}$ is the treatment indicator, $p_b$ is the block treatment assignment rate, $n_b$ is the number of persons in block $b$, $G_{ibk}$ is the subgroup indicator, $n_{bk}$ is the size of subgroup $k$, $\pi_{bk} = n_{bk}/n_b$ is the subgroup share, and $Y_{ib}(t)$ is the potential outcome. Further, we define $S_{ib}$ as a 1/0 indicator of block membership, and $q_b = n_b/n$ as the block population share. We assume SUTVA and complete randomization within each block, where vectors of possible treatment assignments are mutually independent across blocks.

With this notation, we can now define the ATE estimand for blocks containing members of subgroup $k$ as, $\tau_{bk} = \bar{Y}_{bk}(1) - \bar{Y}_{bk}(0)$, and the pooled ATE estimand across such blocks as,

$$\tau_k = \frac{\sum_{b:\pi_{bk}>0}^{h} w_{bk}\tau_{bk}}{\sum_{b:\pi_{bk}>0}^{h} w_{bk}}, \quad (13)$$

where $w_{bk}$ is the block weight, which can differ across subgroups. We set $w_{bk} = n_{bk}$, but other options exist [39]. We allow $n_{bk} = n_b$ and $n_{bk} = 0$.

Consider OLS estimation of the following extension of (6) to blocked RCTs:

$$y_{ib} = \sum_{k=1}^{K} \sum_{b:n_{bk}>0}^{h} \tau_{bk} S_{ib} G_{ibk} \tilde{T}_{ib} + \sum_{k=1}^{K} \sum_{b:n_{bk}>0}^{h} \alpha_{bk} S_{ib} G_{ibk} + \tilde{\mathbf{x}}_{ibk}\boldsymbol{\beta} + \eta_{ib}, \quad (14)$$

where $\tilde{T}_{ib} = T_{ib} - p_b$ and $\tilde{\mathbf{x}}_{ibk} = \mathbf{x}_{ib} - \bar{\mathbf{x}}_{bk}$ are block-centered variables; $\bar{\mathbf{x}}_{bk} = \frac{1}{n_{bk}}\sum_{i=1}^{n_b} G_{ibk}\mathbf{x}_{ib}$ are covariate means; and $\eta_{ib}$ is the error term. The OLS ATE estimator for $\tau_{bk}$ in (14) is,

$$\hat{\tau}_{bk}^x = (\bar{y}_{bk}^1 - \bar{y}_{bk}^0) - (\bar{\mathbf{x}}_{bk}^1 - \bar{\mathbf{x}}_{bk}^0)\hat{\boldsymbol{\beta}}, \quad (15)$$

where $\bar{y}_{bk}^t$ and $\bar{\mathbf{x}}_{bk}^t$ are observed treatment and control group means.

Because a mini-experiment is conducted in each block, we can apply Theorem 1 to $\hat{\tau}_{bk}^x$ as $n \to \infty$ for fixed $h$ (full sample asymptotics as $h \to \infty$ are considered in [40]). To apply the theorem, we redefine the residual as, $R_{ibk}(t) = \frac{G_{ibk}}{\pi_{bk}}(Y_{ib}(t) - \bar{Y}_{bk}(t) - (\mathbf{x}_{ib} - \bar{\mathbf{x}}_{bk})\boldsymbol{\beta})$, invoke (*C1*)-(*C4*) and (*C6*)-(*C8*) for each included block, and add two conditions:

(*C4a*) The block shares, $n_b/n \to q_b^*$ as $n \to \infty$, where $q_b^* > 0$ and $\sum_{b=1}^h q_b^* = 1$.

(*C5a*) The subgroup shares, $n_{bk}/n_b \to \pi_{bk}^*$ as $n \to \infty$, with $0 \le \pi_{bk}^* \le 1$ and $\sum_{k=1}^K \pi_{bk}^* = 1$.

Condition (*C4a*) allows $n_b$ to grow with $n$, and (*C5a*) amends (*C5*) so that $\pi_{bk}$ can equal 0 or 1.

With these conditions, we have the following CLT result: $\frac{1}{\sqrt{\text{Var}(\bar{D}_{bk})}}(\hat{\tau}_{bk}^x - \tau_{bk}) \xrightarrow{d} N(0,1)$, where $\text{Var}(\bar{D}_{bk})$ is defined as in (9) at the block level. The proof parallels the one for Theorem 1, applied to each block. A variance estimator, $\widehat{\text{Var}}(\bar{D}_{bk})$, can be obtained using (12), where $s_{R_{bk}}^2(t)$ is calculated using residuals from the fitted model in (14). The *df* adjustments are $(n_{bk}^1 - Vq_b p_b \pi_{bk}^1 - 1)$ for $s_{R_{bk}}^2(1)$ and $(n_{bk}^0 - Vq_b(1-p_b)\pi_{bk}^0 - 1)$ for $s_{R_{bk}}^2(0)$.

A corollary is that the pooled subgroup estimator across blocks, $\hat{\tau}_k^x = \frac{1}{n_k}\sum_{b:n_{bk}>0}^h n_{bk}\hat{\tau}_{bk}^x$, is consistent and asymptotically normal: $\frac{1}{\sqrt{\text{Var}(\bar{D}_k)}}(\hat{\tau}_k^x - \tau_k) \xrightarrow{d} N(0,1)$, where $\text{Var}(\bar{D}_k) = \frac{1}{(n_k)^2}\sum_{b:n_{bk}>0}^h n_{bk}^2 \text{Var}(\bar{D}_{bk})$. This follows because the $\hat{\tau}_{bk}^x$ estimators are asymptotically independent across blocks, which can be shown using the same arguments as for Corollary 1 in Supplement A. We can estimate $\text{Var}(\bar{D}_k)$ using $\widehat{\text{Var}}(\bar{D}_{bk})$ for each included block. Hypothesis testing for $\hat{\tau}_k^x$ can be conducted using t-tests with $df = (n_k - V\pi_k - 2h)$ or z-tests.

Finally, a future research topic is to develop a CLT for a restricted model that controls for block main effects but excludes block-by-treatment interactions. An example of such a model is to replace the first set of interactions in (14) with $\sum_{k=1}^{K} \tau_{k,R} G_{ibk} \tilde{T}_{ib}$. In this case, the OLS ATE estimator for subgroup $k$ is $\hat{\tau}_{k,R} = \frac{1}{\sum_b w_{bk,R}} \sum_b w_{bk,R} \hat{\tau}_{bk}^x$, where $w_{bk,R} = n_{bk} p_{bk}^1 (1 - p_{bk}^1)$ and $p_{bk}^1 = n_{bk}^1 / n_{bk}$. Thus, $\hat{\tau}_{k,R}$ uses a form of precision weighting to weight the block-specific estimators. It is biased but uses fewer parameters. Full sample CLTs for this estimator are considered in [14] and [21], which become more complex in the subgroup context.

### 4.2. Clustered RCTs

In clustered RCTs, groups rather than individuals are the unit of randomization. Consider a clustered, non-blocked RCT with $m$ total clusters, where $m^1 = mp$ are assigned to the treatment group and $m^0 = m(1-p)$ are assigned to the control group. All persons in the same cluster have the same treatment assignment. Let $m_k$ denote the number of clusters in subgroup $k$, where $m_k^1$ and $m_k^0$ are observed counts. We assume individual-level data are available for analysis, although our results also pertain to data averaged to the cluster level.

We index clusters by $j$. Thus, we have that $T_j = 1$ for treatment clusters and 0 for control clusters, $n_{jk}$ is the number of subgroup $k$ members in cluster $j$, $Y_{ij}(t)$ is the potential outcome for person $i$ in cluster $j$, and so on. We also assume SUTVA and complete randomization as generalized to clustered RCTs [14].

Consider an individual-level subgroup ($G_{ijk} = 1$) where $\pi_{jk} > 0$ for all $j$. In this case, $m_k^1 = m^1$ and $m_k^0 = m^0$ are fixed, and the ATE estimand for subgroup $k$ under the clustered RCT is,

$$\tau_k = \frac{\sum_{j=1}^{m} \sum_{i=1}^{n_j} G_{ijk}(Y_{ij}(1) - Y_{ij}(0))}{n_k} = \frac{\sum_{j=1}^{m} n_{jk}(\bar{Y}_{jk}(1) - \bar{Y}_{jk}(0))}{n_k} = \bar{\bar{Y}}_k(1) - \bar{\bar{Y}}_k(0), \quad (16)$$

where $\bar{Y}_{jk}(t) = \frac{1}{n_{jk}} \sum_{i=1}^{n_j} G_{ijk} Y_{ij}(t)$ is the mean cluster-level outcome, and $\bar{\bar{Y}}_k(t)$ is the grand mean. Here, clusters are weighted by their subgroup sizes, $w_{jk} = n_{jk}$, but other options exist, such as weighting clusters equally to estimate subgroup effects per cluster rather than per person.

Applying OLS to (6) with clustered data yields the following subgroup ATE estimator:

$$\hat{\tau}^x_{k,clus} = (\bar{\bar{y}}^1_k - \bar{\bar{y}}^0_k) - (\bar{\bar{x}}^1_k - \bar{\bar{x}}^0_k)\hat{\boldsymbol{\beta}}, \tag{17}$$

where $\bar{\bar{y}}^t_k = \frac{1}{n^t_k} \sum_{j:T_j=t} \sum_{i=1}^{n_j} G_{ijk} Y_{ij}(t)$ is the mean observed outcome, and similarly for $\bar{\bar{x}}^t_k$.

We see that (17) is a ratio estimator because $n^1_k$ and $n^0_k$ are random under the clustered design (if clusters are weighted unequally). However, this is also the case for the *full sample* estimator, because $n^1$ and $n^0$ (i.e., the summed weights) are also random. Thus, as $m \to \infty$, the full sample CLT results in Schochet et al. [14] for the clustered (and blocked) RCT can be applied to $\hat{\tau}^x_{k,clus}$. This approach is outlined in Supplement C.2 along with a consistent variance estimator using a version of (12) based on estimated cluster-level residuals. Parallel to the HW analysis, Supplement C.2 shows this variance estimator is asymptotically equivalent to the cluster-robust standard error estimator [41].

Finally, Supplement C.3 outlines cluster RCT results for a subgroup analysis defined by a cluster-level characteristic ($G_{jk} = 1$), such as a school performance category, rather than an individual-level characteristic. In this setting, $m^1_k$ and $m^0_k$ become random, which parallels the subgroup analysis for the non-clustered RCT. Note that a similar formulation also applies for the individual-level subgroup analysis when $\pi_{jk} = 0$ for some clusters.

## 5. Simulation analysis

We conducted simulations to examine the finite sample statistical properties of our design-based subgroup ATE estimators. The focus is on the non-clustered RCT because prior full sample simulation results for the clustered RCT also pertain to individual-level subgroup analyses [42,14], as discussed above. For the simulations, we applied the variance estimator in (12) using expected and actual subgroup sizes, for models with and without covariates. We set $\phi_k = 1$ for most specifications, but also adjusted for $\phi_k$ for some runs. We also ran simulations using the HW estimator and several variants of (12) (see Supplement D).

*5.1. Simulation setup*

The following model was used to generate potential outcomes for $K = 2$ subgroups and $V = 2$ pre-treatment covariates:

$$Y_i(0) = G_{i1} + 2G_{i2} + .4G_{i1}x_{i1} + .8G_{i1}x_{i2} + .7G_{i2}x_{i1} + .5G_{i2}x_{i2} + e_i$$
$$Y_i(1) = Y_i(0) + G_{i1}\theta_{i1} + G_{i2}\theta_{i2},$$

(18)

where $e_i$ are *iid* $N(0,1)$ random errors; $x_{i1}$ and $x_{i2}$ are *iid* $N(0,1)$ covariates; and $\theta_{i1}$ and $\theta_{i2}$ are *iid* $N(0,.5)$ and $N(0,.4)$ random errors that capture treatment effect heterogeneity.

We generated 5 draws of potential outcomes using (18) to help guard against unusual draws and report average results. For each draw, we conducted 10,000 replications, randomly assigning units to either the treatment or control group using $p = .5$ (or $p = .4$ or $.6$ for some runs), and only kept randomizations that met our minimum subgroup size criteria for variance estimation. For each replication, we estimated the model in (6) and stored the results. We ran simulations for total sample sizes of $n = 40$, 100, and 200 and Subgroup 1 shares of $\pi_1 = .25, .50,$ and $.75$. To allow for skewed distributions, we also generated model errors and covariates for selected runs using a chi-squared distribution with the same means and variances as above.

In Supplement D, we discuss variants of (12) used in our simulations. These include applying the *df* correction for hypothesis testing in Bell and McCaffrey [43]; subtracting a lower bound on the $\frac{1}{n_k}\Omega^2(\tau_k)$ heterogeneity term; multiplying by $(1 - R^2_{TXk})^{-1}$, where $R^2_{TXk}$ is the $R^2$ from a regression of $G_{ik}\tilde{T}_i$ on $\tilde{\mathbf{x}}_{ik}$ and the other terms in (6); and using the finite sample variance in (11).

*5.2. Simulation results*

Table 1 and Supplement Tables D.1 to D.4 present the simulation results. Of the 300,000 draws of $n_k^1$ and $n_k^0$ used in Table 1, all yielded values of $n_k^1 > 0$ and $n_k^0 > 0$, so these restrictions have little effect on our theory. Focusing on Subgroup 1, we find negligible biases for all specifications with and without baseline covariates. Confidence interval coverage is close to 95 percent using t-distribution cutoff values, even with relatively small subgroup samples, but with slight over-coverage across specifications. Accordingly, Type 1 errors tend to be slightly below the nominal 5 percent level (Tables 1 and D.1). It is interesting that these results differ from those found for the clustered RCT where Type 1 errors tend to be inflated [42,14].

Estimated standard errors (SEs) are close to "true" values, as measured by the standard deviation of the ATE estimates across replications. Consistent with the theory on SE ratios in Section 3.3, the SEs are slightly larger using actual subgroup sizes than expected ones, leading to narrower confidence interval coverage using the expected sizes. Also consistent with the theory, the SEs are slightly smaller for the HW estimator for the model without covariates, and for specifications that adjust for $\phi_k < 1$. Type 1 errors for F-tests to gauge differences in Subgroup 1 and 2 effects are close to 5 percent but tend to be somewhat liberal (Tables D.1-D.2). We find similar results using data generated from a chi-squared distribution (Table D.3) and using $p = .4$ or $.6$ (Table D.4). Finally, applying variants of the variance formula in (12) as detailed in Supplement D does not change the overall findings or improve performance (Table D.3).

**6. Empirical application using the motivating NYC voucher experiment**

To demonstrate our design-based subgroup ATE estimators, we used baseline and outcome data from the NYC School Choice Scholarships Foundation Program (SCSF) [9]. SCSF was funded by philanthropists to provide scholarships to public school students in grades K-4 from low-income families to attend any participating NYC private school. In spring 1997, more than 20,000 students applied to receive a voucher. SCSF then used random lotteries to offer 3-year vouchers of up to $1,400 annually to 1,000 eligible families in the treatment group. Of the remaining families not offered the voucher, 960 were randomly selected to the control group.

SCSF assisted the treatment group in finding private-school placements. More than 78 percent of treatment families used a voucher, for 2.6 years on average, where 98 percent of users attended parochial schools. Here, we focus on estimating ATEs (i.e., intention-to-treat effects on the voucher offer) for two race/ethnicity subgroups as defined in [9]: African Americans and Latinos who each comprise about 47 percent of the sample. The study authors hypothesized that African Americans might benefit more from the vouchers as they tended to live in more disadvantaged communities with lower-performing public schools.

Following the original study [9], the primary outcomes for our analysis are composite national percentile rankings in math and reading from the study-administered Iowa Test of Basic Skills. We focus on first follow-up year test scores, where the response rate was 78 percent for treatments and 71 percent for controls. Our goal is not to replicate study results but to illustrate our subgroup ATE estimators.

The voucher study was a blocked RCT. Applicants from schools with average test scores below the city median were assigned a higher probability of winning a scholarship, and blocks were also formed by lottery date and family size (with 30 blocks in total). The design is also partly clustered because families were randomized, where all eligible children within a family could receive a scholarship; 30 percent of families had at least two children in the evaluation.

We used (14) for ATE estimation and (12) for variance estimation for each block, where blocks were weighted by their subgroup sizes to obtain overall subgroup effects. To adjust for clustering, we averaged data to the family level. Following [9], we used weights to adjust for missing follow-up test scores. We ran models without covariates and those that included baseline ITBS scores to increase precision, though they were not collected for the entire kindergarten cohort. Following the original study, other demographic covariates were not included in the models due to the large number of blocks.

Table 2 presents the subgroup findings that mirror those from the original study. We find that the offer of a voucher had no effect on test scores overall or for Latinos across specifications. The effects on African Americans are also not statistically significant at the 5 percent level for the model without baseline test scores. However, these effects become positive and statistically significant for the model with baseline scores, that excludes the kindergarteners but nonetheless yields standard errors that are reduced by about 12 percent. These effects are 4.7 percentile ranking points, which translates into a .26 standard deviation increase, with a significant F-test for the subgroup interaction effect ($p$-value = .028). The effects for African Americans remain significant using the sample with baseline test scores without controlling for them in the model.

We find across specifications that the design-based standard errors are nearly identical using actual and expected sample sizes. Further, consistent with theory, the design-based standard errors are slightly larger than the HW standard errors, but both yield the same study conclusions: the vouchers did not improve test scores overall, but there is evidence they had a positive effect on African American students in grades 1-4. A detailed reanalysis of the original study data, however, cautions that the results for African Americans are sensitive to alternative race/ethnicity definitions and should be interpreted carefully [10].

## 7. Conclusions

This article considered design-based RCT methods for ATE estimation for discrete subgroups defined by pre-treatment sample characteristics. Our subgroup estimators derive from the Neyman-Rubin-Holland model that underlies experiments and are based on simple least squares regression methods. We considered ratio estimators due to the randomness of subgroup sample sizes in the treatment and control groups. The design-based approach is appealing in that it applies to continuous, binary, and discrete outcomes, and is nonparametric in that makes no assumptions about the distribution of potential outcomes or the model functional form.

We developed a finite population CLT for our subgroup ATE estimators under the non-clustered RCT, allowing for baseline covariates to improve precision. The main difference between our design-based CLT and prior full sample ones is that the asymptotic variance for the subgroup estimator is based on expected subgroup sizes rather than actual ones. Another difference is that the subgroup variance includes a finite sample adjustment ($\phi_k$) that reflects the single treatment indicator shared by the subgroups. To apply the estimators in practice, we discussed simple consistent variance estimators using regression residuals that are asymptotically equivalent to robust estimators, but with finite sample degrees-of-freedom adjustments that derive directly from the experimental design. Our re-analysis of the NYC Voucher experiment demonstrated the simplicity of the methods, while maintaining statistical rigor.

A contribution of this work is that it provides a unified design-based framework for subgroup analyses across a range of RCT designs. We discussed extensions of our asymptotic theory to blocked and clustered designs. We also discussed extensions to other commonly used estimators with random treatment-control sample sizes (or weights): post-stratification estimators that average subgroup estimators to obtain overall effects, weighted estimators to adjust for data nonresponse, and estimators from Bernoulli trials.

Our simulations for the non-clustered RCT show that the subgroup ATE estimators yield low bias and confidence interval coverage near nominal levels, although with slight over-coverage. This is somewhat surprising as the simulation literature on design-based and robust variance estimators for clustered RCTs—that also applies to the individual-level subgroup context—shows the opposite issue of under-coverage [42,14].

Our simulations find very similar results using either actual or expected subgroup sample sizes for variance estimation. As demonstrated in several ways, this occurs because the difference between the observed subgroup proportions, $\pi_k^1$ and $\pi_k^0$, and their expected value, $\pi_k$, decreases exponentially with the overall sample size. This finding justifies the typical approach of using actual subgroup sample sizes for variance estimation, which blurs the distinction between a subgroup analysis conditional on the observed treatment-control subgroup sizes and an unconditional subgroup analysis that was considered here.

The free *RCT-YES* software ([www.rct-yes.com](www.rct-yes.com)), funded by the U.S. Department of Education, estimates ATEs for both full sample and baseline subgroup analyses using the design-based methods discussed in this article using either R or Stata. The software applies actual sample sizes for the variance formulas for subgroup analyses and allows for general weights. The software also allows for multi-armed trials with multiple treatment conditions.

**Data Availability Statement**. The NYC Voucher data for the empirical analysis were obtained under a restricted data use license agreement with Mathematica. Per license requirements, these data cannot be shared with journal readers. However, to the best of my knowledge, these data can be obtained, and I would be happy to provide the SAS and R programs used for the analysis.

Panel 1A

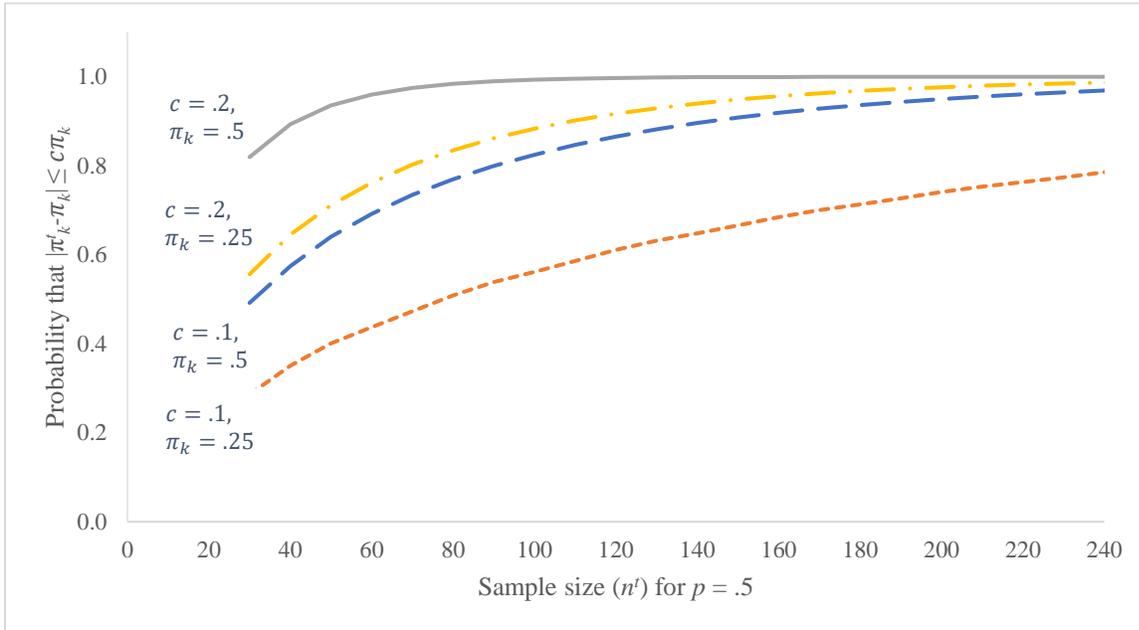

Panel 1B

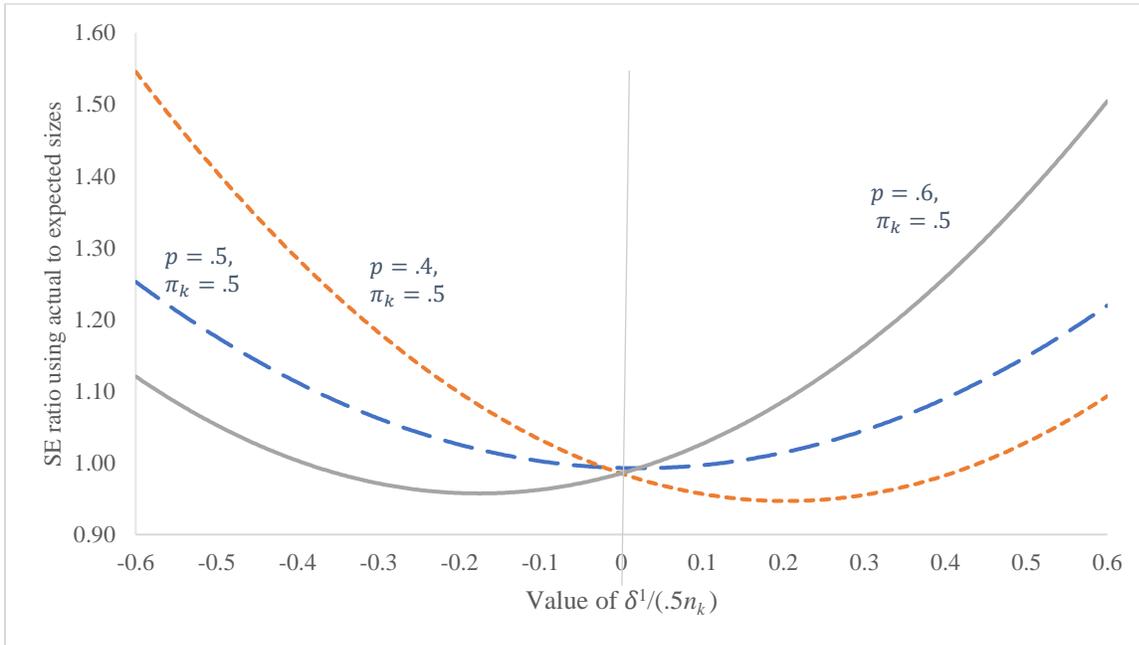

**Figure 1.** Probabilities for the differences, $(\pi_k^t - \pi_k)$, relative to $\pi_k$ (Panel 1A), and standard error ratios using actual to expected subgroup sizes, as a function of $(n_k^1 - n_k p)/.5 n_k$ (Panel 1B).

Notes: See text for definitions and formulas. Panel 1B assumes $n = 100$, $\varphi = 1.1$ and $\vartheta = .05$.
SE = Standard error.

**Table 1.** Simulation results for the subgroup ATE estimators.

| Model specification | Bias of ATE estimator[a] | Confidence interval coverage | True standard error[a,b] | Mean estimated standard error |
|---|---|---|---|---|
| **Model without covariates** | | | | |
| Sample size: $n = 40, \pi_1 = .50$ | | | | |
| Design-based (DB), actual subgroup sizes, $\phi_k = 1$ | -.002 | .954 | .646 | .640 |
| DB, expected sizes, $\phi_k = 1$ | -.002 | .952 | .646 | .633 |
| DB, actual sizes, adjust for $\phi_k$ | -.002 | .948 | .646 | .621 |
| Huber-White (HW) | -.002 | .953 | .646 | .639 |
| Sample size: $n = 100, \pi_1 = .50$ | | | | |
| DB, actual sizes, $\phi_k = 1$ | .000 | .958 | .376 | .385 |
| DB, expected sizes, $\phi_k = 1$ | .000 | .957 | .376 | .383 |
| DB, actual sizes, adjust for $\phi_k$ | .000 | .956 | .376 | .381 |
| HW | .000 | .958 | .376 | .385 |
| Sample size: $n = 100, \pi_1 = .25$ | | | | |
| DB, actual sizes, $\phi_k = 1$ | .000 | .953 | .626 | .628 |
| DB, expected sizes, $\phi_k = 1$ | .000 | .951 | .626 | .618 |
| DB, actual sizes, adjust for $\phi_k$ | .000 | .948 | .626 | .613 |
| HW | .000 | .948 | .626 | .613 |
| **Model with two covariates** | | | | |
| Sample size: $n = 40, \pi_1 = .50$ | | | | |
| DB, actual sizes, $\phi_k = 1$ | .002 | .950 | .501 | .482 |
| DB, expected sizes, $\phi_k = 1$ | .002 | .948 | .501 | .476 |
| DB, actual sizes, adjust for $\phi_k$ | .002 | .944 | .501 | .467 |
| HW | .002 | .953 | .501 | .494 |
| Sample size: $n = 100, \pi_1 = .50$ | | | | |
| DB, actual sizes, $\phi_k = 1$ | .000 | .961 | .298 | .305 |
| DB, expected sizes, $\phi_k = 1$ | .000 | .960 | .298 | .303 |
| DB, actual sizes, adjust for $\phi_k$ | .000 | .959 | .298 | .302 |
| HW | .000 | .962 | .298 | .308 |
| Sample size: $n = 100, \pi_1 = .25$ | | | | |
| DB, actual sizes, $\phi_k = 1$ | .000 | .956 | .486 | .482 |
| DB, expected sizes, $\phi_k = 1$ | .000 | .954 | .486 | .475 |
| DB, actual sizes, adjust for $\phi_k$ | .000 | .951 | .486 | .470 |
| HW | .000 | .952 | .486 | .470 |

Notes. See text for simulation details. The calculations assume two subgroups with a focus on results for Subgroup 1, a treatment assignment rate of $p = .50$, and normally distributed covariates and errors. For each specification, the figures are based on 10,000 simulations for each of 5 potential outcome draws, and the findings average across the 5 draws. Ordinary least square (OLS) methods are used for ATE estimation using the model in (6), and design-based standard errors are obtained using (12). Huber-White estimates are obtained using the lm_robust procedure in R.

ATE = Average treatment effect; DB = Design-based; HW = Hubert-White.

[a] Biases and true standard errors are the same for all specifications within each sample size category because they use the same data and OLS model for ATE estimation.

[b] True standard errors are measured as the standard deviation of the estimated treatment effects across simulations.

**Table 2.** Estimated ATEs on composite test scores for the NYC voucher experiment.

| Model specification | Overall sample | African American | Latino |
|---|---|---|---|
| **Model excludes baseline test scores** | | | |
| Design-based, actual subgroup sizes | 0.25 (1.06) | 2.54 (1.45) | -0.86 (1.58) |
| Design-based, expected subgroup sizes | 0.25 (1.06) | 2.54 (1.45) | -0.86 (1.58) |
| Huber-White | 0.25 (1.03) | 2.54 (1.42) | -0.86 (1.49) |
| Design-based: actual sizes using sample with baseline test scores | 0.88 (1.30) | 4.47* (1.73) | -1.11 (1.76) |
| **Model includes baseline test scores** | | | |
| Design-based, actual subgroup sizes | 1.70 (1.01) | 4.70* (1.27) | 0.50 (1.44) |
| DB, expected subgroup sizes | 1.70 (1.01) | 4.70* (1.27) | 0.50 (1.44) |
| Huber-White | 1.70 (0.98) | 4.70* (1.24) | 0.50 (1.39) |
| Student sample size (without / with baseline test scores) | 2,012 / 1,434 | 902 / 643 | 964 / 682 |

Notes: Standard errors are in parentheses. See text for ATE and standard error formulas. All estimates are weighted to adjust for follow-up test score nonresponse.

ATE = Average treatment effect.

* Statistically significant at the 5 percent level, two-tailed test.

# Supplementary Materials for "Design-Based RCT Estimators and Central Limit Theorems for Baseline Subgroup and Related Analyses"

## A. Proof of Theorem 1

To prove Theorem 1, we first prove consistency and then the central limit theorem (CLT). We also prove a corollary to Theorem 1 discussed in Remark 5 in the main text. We use definitions and notation from the main text. Note that the proof also applies to the model without covariates by setting $\boldsymbol{\beta} = \mathbf{0}$. Our approach follows Schochet et al. [1] and uses results in Li and Ding [2] and Scott and Wu [3].

### A.1. Proof of consistency

Our goal is to show that $\hat{\tau}_k^x \xrightarrow{p} \mu_{Yk}(1) - \mu_{Yk}(0)$ as $n \to \infty$, where $\mu_{Yk}(1)$ and $\mu_{Yk}(0)$ are finite limiting values of $\bar{Y}_k(1)$ and $\bar{Y}_k(0)$, the mean potential outcomes for subgroup $k$ in the treatment and control conditions.

Consider ordinary least squares (OLS) estimation of the model in (6) in the main text, where $\tilde{\mathbf{z}}_{ik} = (G_{i1}\tilde{T}_i, \ldots, G_{iK}\tilde{T}_i, G_{i1}, \ldots, G_{iK}, \tilde{\mathbf{x}}_{ik})$ is the vector of model explanatory variables with parameter vector, $\boldsymbol{\gamma}$. The OLS estimator using data on the full sample is,

$$\hat{\boldsymbol{\gamma}} = \begin{pmatrix} \hat{\tau}_1^x \\ \vdots \\ \hat{\tau}_K^x \\ \hat{\alpha}_1^x \\ \vdots \\ \hat{\alpha}_K^x \\ \hat{\boldsymbol{\beta}} \end{pmatrix} = (\frac{1}{n}\sum_{i=1}^n \tilde{\mathbf{z}}'_{ik}\tilde{\mathbf{z}}_{ik})^{-1} \frac{1}{n}\sum_{i=1}^n \tilde{\mathbf{z}}'_{ik} y_i, \qquad (A.1)$$

where $y_i = T_i Y_i(1) + (1 - T_i) Y_i(0)$ is the observed outcome. Expanding this estimator yields,

$$\begin{pmatrix} \hat{\tau}_1^x \\ \vdots \\ \hat{\tau}_K^x \\ \hat{\alpha}_1^x \\ \vdots \\ \hat{\alpha}_K^x \\ \hat{\boldsymbol{\beta}} \end{pmatrix} = \begin{bmatrix} \frac{1}{n}\sum_i G_{i1}\tilde{T}_i^2 & \cdots & 0 & \frac{1}{n}\sum_i G_{i1}\tilde{T}_i & \cdots & 0 & \frac{1}{n}\sum_i G_{i1}\tilde{T}_i \tilde{\mathbf{x}}_{i1} \\ \vdots & \ddots & \vdots & \vdots & \ddots & \vdots & \vdots \\ 0 & \cdots & \frac{1}{n}\sum_i G_{iK}\tilde{T}_i^2 & 0 & \cdots & \frac{1}{n}\sum_i G_{iK}\tilde{T}_i & \frac{1}{n}\sum_i G_{iK}\tilde{T}_i \tilde{\mathbf{x}}_{iK} \\ \frac{1}{n}\sum_i G_{i1}\tilde{T}_i & \cdots & 0 & \frac{1}{n}\sum_i G_{i1} & \cdots & 0 & \mathbf{0}_{Vx1} \\ \vdots & \ddots & \vdots & \vdots & \ddots & \vdots & \vdots \\ 0 & \cdots & \frac{1}{n}\sum_i G_{iK}\tilde{T}_i & 0 & \cdots & \frac{1}{n}\sum_i G_{iK} & \mathbf{0}_{Vx1} \\ \frac{1}{n}\sum_i G_{i1}\tilde{T}_i \tilde{\mathbf{x}}'_{i1} & \cdots & \frac{1}{n}\sum_i G_{iK}\tilde{T}_i \tilde{\mathbf{x}}'_{iK} & \mathbf{0}_{1xV} & \cdots & \mathbf{0}_{1xV} & \frac{1}{n}\sum_k \sum_i G_{ik}\tilde{\mathbf{x}}'_{ik}\tilde{\mathbf{x}}_{ik} \end{bmatrix}^{-1} \qquad (A.2)$$

$$\begin{bmatrix} \frac{1}{n}\sum_i G_{i1}\tilde{T}_i y_i \\ \vdots \\ \frac{1}{n}\sum_i G_{iK}\tilde{T}_i y_i \\ \frac{1}{n}\sum_i G_{i1} y_i \\ \vdots \\ \frac{1}{n}\sum_i G_{iK} y_i \\ \frac{1}{n}\sum_k \sum_i G_{ik}\tilde{\mathbf{x}}_{ik} y_i \end{bmatrix}.$$

Due to the centering of the variables, $\frac{1}{n}\sum_{i=1}^n \tilde{\mathbf{z}}'_{ik}\tilde{\mathbf{z}}_{ik}$ converges to a block diagonal matrix as $n \to \infty$. This can be seen by examining, in turn, each of the non-zero off-diagonal terms in (A.2):

$$\frac{1}{n}\sum_i G_{ik}\tilde{T}_i = p(1-p)[\pi_k^1 - \pi_k^0] \xrightarrow{p} p^*(1-p^*)[\pi_k^* - \pi_k^*] = 0 \tag{A.3a}$$

and

$$\frac{1}{n}\sum_i G_{i1}\tilde{T}_i\tilde{\mathbf{x}}_i = p(1-p)[\pi_k^1\bar{\mathbf{x}}_k^1 - \pi_k^0\bar{\mathbf{x}}_k^0] \xrightarrow{p} p^*(1-p^*)[\pi_k^*\boldsymbol{\mu}_X - \pi_k^*\boldsymbol{\mu}_X] = 0, \tag{A.3b}$$

where $\boldsymbol{\mu}_X$ is the limiting value of $\bar{\mathbf{x}}_k$, assumed finite. These two results rely on Conditions (*C4*)-(*C6*) in Theorem 1 and the continuous mapping theorem. Further, for (A.3b), we invoke the weak law of large numbers for each covariate using Theorem B in Scott and Wu [3], which applies because the covariate variances, $S^2_{x_k,v}$, are assumed to have finite limits by (*C8*). Thus, $\bar{\mathbf{x}}_k^t - \bar{\mathbf{x}}_k \xrightarrow{p} \mathbf{0}$, which implies that $\bar{\mathbf{x}}_k^t - \boldsymbol{\mu}_X = (\bar{\mathbf{x}}_k^t - \bar{\mathbf{x}}_k) - (\bar{\mathbf{x}}_k - \boldsymbol{\mu}_X) \xrightarrow{p} \mathbf{0}$, so $\bar{\mathbf{x}}_k^t \xrightarrow{p} \boldsymbol{\mu}_X$.

Since the design matrix is block diagonal in the limit, it follows that $\hat{\tau}_k^x$ has the same limiting value as $\sum_{i=1}^n G_{ik}\tilde{T}_i y_i / \sum_{i=1}^n G_{ik}\tilde{T}_i^2$. To calculate this limit, we use the following results:

$$\frac{1}{n}\sum_{i=1}^n G_{ik}\tilde{T}_i^2 = p(1-p)[(1-p)\pi_k^1 + p\pi_k^0] \xrightarrow{p} p^*(1-p^*)\pi_k^* \tag{A.4a}$$

and

$$\frac{1}{n}\sum_{i=1}^n G_{ik}\tilde{T}_i y_i = p(1-p)[\pi_k^1 \bar{y}_k^1 - \pi_k^0 \bar{y}_k^0] \xrightarrow{p} p^*(1-p^*)\pi_k^*[\mu_{Yk}(1) - \mu_{Yk}(0)]. \tag{A.4b}$$

We have that (A.4b) holds using a similar argument as for (A.3b), where $\bar{y}_k^t \xrightarrow{p} \mu_{Yk}(t)$ as $\bar{y}_k^t - \bar{Y}_k(t) \xrightarrow{p} 0$ due to the assumed finite variances for the potential outcomes (so we can again apply Theorem B in [3]), and noting that, $\bar{y}_k^t - \mu_{Yk}(t) = (\bar{y}_k^t - \bar{Y}_k(t)) - (\bar{Y}_k(t) - \mu_{Yk}(t)) \xrightarrow{p} 0$. If we now divide (A.3b) by (A.3a), we find that,

$$\hat{\tau}_k^x \xrightarrow{p} \mu_{Yk}(1) - \mu_{Yk}(0), \tag{A.5}$$

which establishes consistency.

For the CLT proof, we will also need asymptotic values for the estimated subgroup intercepts, $\hat{\alpha}_k^x$, and covariate parameters, $\widehat{\boldsymbol{\beta}}$. First, note that $\hat{\alpha}_k^x$ has the same limiting value as $\frac{1}{\Sigma_i G_{ik}} \Sigma_i G_{ik} y_i$, so we can use a similar argument as for $\hat{\tau}_k^x$ to show that,

$$\hat{\alpha}_k^x \xrightarrow{p} p^* \mu_{Yk}(1) + (1-p^*) \mu_{Yk}(0) = \alpha^*, \tag{A.6}$$

which is the mean potential outcome for subgroup $k$.

Similarly, the OLS estimator for the covariates, $\widehat{\boldsymbol{\beta}}$, has the same limiting value as $\left[\frac{1}{n}\sum_{k=1}^K \sum_{i=1}^n G_{ik} \tilde{\mathbf{x}}'_{ik} \tilde{\mathbf{x}}_{ik}\right]^{-1} \left[\frac{1}{n}\sum_{k=1}^K \sum_{i=1}^n G_{ik} \tilde{\mathbf{x}}'_{ik} y_i\right]$. For the first bracketed term, we have that,

$$\frac{1}{n} \sum_{k=1}^K \sum_{i=1}^n G_{ik} \tilde{\mathbf{x}}'_{ik} \tilde{\mathbf{x}}_{ik} = \sum_{k=1}^K \frac{n_k}{n} \mathbf{S}_{\mathbf{x},k}^2 \xrightarrow{p} \sum_{k=1}^K \pi_k^* \boldsymbol{\Omega}_{\mathbf{x},k}^2, \tag{A.7}$$

where $\boldsymbol{\Omega}_{\mathbf{x},k}^2$ is the finite limiting value of $\mathbf{S}_{\mathbf{x},k}^2$ (see $(C8)$).

For the second bracketed term, if we insert the relation, $y_i = T_i Y_i(1) + (1-T_i) Y_i(0)$, we find that,

$$\frac{1}{n} \sum_{k=1}^K \sum_{i=1}^n G_{ik} \tilde{\mathbf{x}}'_{ik} y_i = \frac{1}{n} \sum_{k=1}^K \sum_{i=1}^n T_i G_{ik} \tilde{\mathbf{x}}'_{ik} Y_i(1) + \frac{1}{n} \sum_{k=1}^K \sum_{i=1}^n (1-T_i) G_{ik} \tilde{\mathbf{x}}'_{ik} Y_i(0). \tag{A.8}$$

We can now invoke the weak law of large numbers for each term on the right-hand side of (A.8), again using Theorem B in [3], which applies because the $\mathbf{S}_{\mathbf{x}Y,k}^2(t)$ covariances are assumed to have finite limits under $(C8)$. We find then that the first term has the same asymptotic value as $\frac{n^1}{n} \sum_{k=1}^K \frac{n_k}{n} \mathbf{S}_{\mathbf{x},Y,k}^2(1)$ and the second term has the same asymptotic value as $\frac{n^0}{n} \sum_{k=1}^K \frac{n_k}{n} \mathbf{S}_{\mathbf{x},Y,k}^2(0)$. If $\boldsymbol{\Omega}_{\mathbf{x},Y,k}^2(t)$ is the finite limiting value of $\mathbf{S}_{\mathbf{x},Y,k}^2(t)$, then after some algebra we find that,

$$\frac{1}{n} \sum_{k=1}^K \sum_{i=1}^n G_{ik} \tilde{\mathbf{x}}'_{ik} y_i \xrightarrow{p} \sum_{k=1}^K \pi_k^* \left[p^* \boldsymbol{\Omega}_{\mathbf{x},Y,k}^2(1) + (1-p^*) \boldsymbol{\Omega}_{\mathbf{x},Y,k}^2(0)\right]. \tag{A.9}$$

Finally, putting together the pieces in (A.8) and (A.9), we find that as $n \to \infty$,

$$\widehat{\boldsymbol{\beta}} \xrightarrow{p} \left(\sum_{k=1}^K \pi_k \boldsymbol{\Omega}_{\mathbf{x},k}^2\right)^{-1} \left[\sum_{k=1}^K p^* \pi_k^* \boldsymbol{\Omega}_{\mathbf{x},Y,k}^2(1) + (1-p^*) \pi_k^* \boldsymbol{\Omega}_{\mathbf{x},Y,k}^2(0)\right] = \boldsymbol{\beta}^*, \tag{A.10}$$

which are regression coefficients that would be obtained from a weighted regression of $\hat{\alpha}_k$ (the mean potential outcomes) on the covariates using $\pi_k^*$ as weights. ∎

## A.2. Proof of CLT

The proof of the CLT in Theorem 1 involves three steps. First, we prove a CLT for a subgroup mean for either the treatment or control group, assuming $\boldsymbol{\beta}$ is known. Second, we prove a CLT for the treatment-control difference in these subgroup means. Finally, we prove the CLT in Theorem 1 based on the estimated OLS coefficients, $\widehat{\boldsymbol{\beta}}$.

### A.2.1. CLT for a single subgroup mean assuming $\boldsymbol{\beta}$ is known

Before presenting our lemma, we require several definitions. First, for $t \in \{1,0\}$, recall from the main text that $R_{ik}(t) = \frac{G_{ik}}{\pi_k}(Y_i(t) - \bar{Y}_k(t) - (\mathbf{x}_i - \bar{\mathbf{x}}_k)\boldsymbol{\beta})$ is the residualized, mean-centered potential outcome for subgroup $k$ in the treatment or control condition, with variance $S_{R_k}^2(t) = \frac{1}{n-1}\sum_{i=1}^{n} R_{ik}^2(t)$. Here, we assume $\boldsymbol{\beta} = \left(\sum_{k=1}^{K} \pi_k \mathbf{S}_{\mathbf{x},k}^2\right)^{-1}\left[\sum_{k=1}^{K} p\pi_k \mathbf{S}_{\mathbf{x},Y,k}^2(1) + \sum_{k=1}^{K}(1-p)\pi_k \mathbf{S}_{\mathbf{x},Y,k}^2(0)\right]$ is known. Note that $\bar{R}_k(t) = \frac{1}{n}\sum_{i=1}^{n} R_{ik}(t) = 0$.

Second, let $\bar{r}_k(t)$ denote the mean of $R_{ik}(t)$ averaged over research group $t$:

$$\bar{r}_k(t) = \frac{1}{n^t} \sum_{i:T_i=t}^{n} R_{ik}(t). \tag{A.11}$$

Finally, recall from the main text that $\bar{y}_k^t = \frac{1}{n^t \pi_k^t}\sum_{i:T_i=t}^{n} G_{ik} Y_i(t)$ is the observed mean outcome; $\bar{Y}_k(t) = \frac{1}{n\pi_k}\sum_{i=1}^{n} G_{ik} Y_i(t)$ is the mean potential outcome in the finite population; and similarly for the covariate means, $\bar{\mathbf{x}}_k^t$ and $\bar{\mathbf{x}}_k$.

Next, we present a CLT lemma for $[\bar{y}_k^t - (\bar{\mathbf{x}}_k^t - \bar{\mathbf{x}}_k)\boldsymbol{\beta}]$ around its finite population mean, $\bar{Y}_k(t)$, assuming $\boldsymbol{\beta}$ is known, where $\bar{\rho}_k(t)$ is the associated covariate-adjusted residual:

$$\bar{\rho}_k(t) = [\bar{y}_k^t - (\bar{\mathbf{x}}_k^t - \bar{\mathbf{x}}_k)\boldsymbol{\beta} - \bar{Y}_k(t)]. \tag{A.12}$$

**Lemma A1.** Under Conditions (*C1*)-(*C8*) of Theorem 1, for $t \in \{1,0\}$, $k \in \{1, \ldots, K\}$, and $K \geq 1$, as $n \to \infty$,

$$\frac{\bar{\rho}_k(t)}{\sqrt{\frac{(1-f^t)}{n^t}S_{R_k}^2(t)}} \xrightarrow{d} N(0,1). \tag{A.13}$$

*Proof.* Under (*C1*)-(*C8*), we have from Theorem 1 in Li and Ding [2] that,

$$\frac{\bar{r}_k(t)}{\sqrt{\frac{(1-f^t)}{n^t}S_{R_k}^2(t)}} \xrightarrow{d} N(0,1),$$

where $f^t = n^t/n$ and $S^2_{R_k}(t)$ have finite limits. Note next that we can express $\bar{r}_k(t)$ as,

$$\bar{r}_k(t) = \frac{1}{n^t} \sum_{i:T_i=t}^{n} R_{ik}(t) = [\bar{y}^t_k - (\bar{\mathbf{x}}^t_k - \bar{\mathbf{x}}_k)\boldsymbol{\beta} - \bar{Y}_k(t)]\frac{\pi^t_k}{\pi_k} = \bar{\rho}_k(t)\frac{\pi^t_k}{\pi_k}. \tag{A.14}$$

Rearranging terms in (A.14), we find that,

$$\bar{\rho}_k(t) = \bar{r}_k(t)\frac{\pi_k}{\pi^t_k}, \tag{A.15}$$

where $\pi_k/\pi^t_k \xrightarrow{p} 1$ due to (C6). Thus, by Slutsky's theorem, we have that the residual, $\bar{\rho}_k$, has the same asymptotic distribution as $\bar{r}_k(t)$, which establishes the lemma. ∎

### A.2.2. CLT for the treatment-control difference in subgroup means assuming β is known

We now prove a CLT for the difference in the covariate-adjusted sample means, $[\bar{y}^1_k - (\bar{\mathbf{x}}^1_k - \bar{\mathbf{x}}_k)\boldsymbol{\beta}]$ and $[\bar{y}^0_k - (\bar{\mathbf{x}}^0_k - \bar{\mathbf{x}}_k)\boldsymbol{\beta}]$ around its finite population ATE estimand, $(\bar{Y}_k(1) - \bar{Y}_k(0))$. Let $\bar{d}_k$ denote the corresponding residual:

$$\bar{d}_k = \bar{\rho}_k(1) - \bar{\rho}_k(0) = (\bar{y}^1_k - \bar{y}^0_k) - (\bar{\mathbf{x}}^1_k - \bar{\mathbf{x}}^0_k)\boldsymbol{\beta} - (\bar{Y}_k(1) - \bar{Y}_k(0)). \tag{A.16}$$

Further, as defined in (8) from the main text, let $\text{Var}(\bar{D}_k)$ be the variance of the mean treatment-control difference in the $R_{ik}(t)$ residuals:

$$\text{Var}(\bar{D}_k) = \frac{S^2_{R_k}(1)}{n^1} + \frac{S^2_{R_k}(0)}{n^0} - \frac{S^2(\tau_k)}{n}, \tag{A.17}$$

where $S^2(\tau_k) = \frac{1}{n-1}\sum_{i=1}^{n}(R_{ik}(1) - R_{ik}(0))^2$ is the heterogeneity of individual-level treatment effects across members of subgroup $k$.

**Lemma A2.** Under (C1)-(C8) of Theorem 1, for $k \in \{1, ..., K\}$ and $K \geq 1$, as $n \to \infty$,

$$\frac{\bar{d}_k}{\sqrt{\text{Var}(\bar{D}_k)}} \xrightarrow{d} N(0,1). \tag{A.18}$$

*Proof.* Under (C1)-(C8), we have from Theorem 4 in Li and Ding [2] that,

$$\frac{\bar{r}_k(1) - \bar{r}_k(0)}{\sqrt{\text{Var}(\bar{D}_k)}} \xrightarrow{d} N(0,1), \tag{A.19}$$

where $\bar{r}_k(t)$ is defined as in (A.11). Next, using (A.15), we can express $(\bar{r}_k(1) - \bar{r}_k(0))$ as,

$$\bar{r}_k(1) - \bar{r}_k(0) = \bar{\rho}_k(1)\frac{\pi_k^1}{\pi_k} - \bar{\rho}_k(0)\frac{\pi_k^0}{\pi_k}. \tag{A.20}$$

From (A.20) and Lemma A1, we have CLTs for $(\bar{r}_k(1) - \bar{r}_k(0))$, $\bar{\rho}_k(1)$, and $\bar{\rho}_k(0)$. Further, under (C6), we have that $\pi_k^t/\pi_k \xrightarrow{p} 1$. Thus, our goal is to use these results to prove a CLT for $(\bar{\rho}_k(1) - \bar{\rho}_k(0))$.

To do this, we follow [1] by using (A.20) to express $(\bar{\rho}_k(1) - \bar{\rho}_k(0))$ as,

$$\frac{\bar{\rho}_k(1) - \bar{\rho}_k(0)}{\sqrt{\text{Var}(\bar{D}_k)}} = \frac{(\bar{r}_k(1) - \bar{r}_k(0))}{\sqrt{\text{Var}(\bar{D}_k)}} + \left(\frac{\bar{\rho}_k(1)}{V(1)}\right)\left(\frac{V(1)}{\sqrt{\text{Var}(\bar{D}_k)}}\right)\left(1 - \frac{\pi_k^1}{\pi_k}\right)$$
$$- \left(\frac{\bar{\rho}_k(0)}{V(0)}\right)\left(\frac{V(0)}{\sqrt{\text{Var}(\bar{D}_k)}}\right)\left(1 - \frac{\pi_k^0}{\pi_k}\right), \tag{A.21}$$

where $V(t) = \sqrt{\frac{(1-f^t)}{n^t}S_{R_k}^2(t)}$. The first term on the right-hand side in (A.21) is asymptotically normal by Theorem 4 in [2]. The second and third terms vanish because (i) the first bracketed terms are each asymptotically normal by Lemma A1 from above, (ii) the second bracketed terms each converge to a constant due to the finite variance assumptions, and (iii) the third bracketed terms each converge in probability to 0 by (C6). Thus, by Slutsky's theorem, $(\bar{\rho}_k(1) - \bar{\rho}_k(0))$ has the same asymptotic distribution as $(\bar{r}_k(1) - \bar{r}_k(0))$, which establishes the lemma. ∎

### A.2.3. Proof of Theorem 1 using the OLS estimator, $\widehat{\boldsymbol{\beta}}$, and Corollary 1 from Remark 7

The previous section provided a CLT when $\boldsymbol{\beta}$ is known. In this section, we prove the CLT from Theorem 1 that is based on the OLS estimator for the covariates, $\widehat{\boldsymbol{\beta}}$. To do this, we express the ATE estimator from (7) in the main text as follows:

$$\hat{\tau}_k^x = \hat{\tau}_k^{x\beta} - (\bar{\mathbf{x}}_k^1 - \bar{\mathbf{x}}_k^0)(\widehat{\boldsymbol{\beta}} - \boldsymbol{\beta}), \tag{A.22}$$

where $\hat{\tau}_k^{x\beta} = (\bar{y}_k^1 - \bar{y}_k^0) - (\bar{\mathbf{x}}_k^1 - \bar{\mathbf{x}}_k^0)\boldsymbol{\beta}$. From Lemma A2 above, we have a CLT for $\hat{\tau}_k^{x\beta}$, so we show next that the difference between $\hat{\tau}_k^x$ and $\hat{\tau}_k^{x\beta}$ is $o_p(n^{-1/2})$, which implies the two estimators have the same asymptotic distribution [1,2].

From Lemma A1, we have a CLT for each element of $\bar{\mathbf{x}}_k^1$ and $\bar{\mathbf{x}}_k^0$. This implies that $[\bar{x}_k^1 - \bar{x}_k]_v = O_p(n^{-1/2})$ and $[\bar{x}_k^0 - \bar{x}_k]_v = O_p(n^{-1/2})$ for each covariate $v \in \{1, \dots, V\}$. Thus, $[\bar{x}_k^1 - \bar{x}_k^0]_v = [\bar{x}_k^1 - \bar{x}_k - (\bar{x}_k^0 - \bar{x}_k)]_v = O_p(n^{-1/2})$. Further, the consistency results from Section A.1 imply that $(\widehat{\boldsymbol{\beta}} - \boldsymbol{\beta}) = o_p(1)$. Thus, $(\bar{\mathbf{x}}_k^1 - \bar{\mathbf{x}}_k^0)(\widehat{\boldsymbol{\beta}} - \boldsymbol{\beta}) = o_p(n^{-1/2})$, which establishes the CLT. ∎

**Corollary 1**. Under the conditions of Theorem 1, as $n \to \infty$, $\hat{\tau}_k^x$ and $\hat{\tau}_{k'}^x$ for subgroups $k$ and $k'$ are asymptotically independent for $(k, k') \in \{1, \ldots, K\}$. Further, the joint asymptotic distribution of the $K$ subgroup ATE estimators, $(\hat{\tau}_1^x, \ldots, \hat{\tau}_K^x)$, is multivariate normal.

*Proof.* In the proof of Theorem 1, we showed that $\hat{\tau}_k^x$ using the estimated $\widehat{\boldsymbol{\beta}}$ has the same asymptotic distribution as $\hat{\tau}_k^{x\beta}$ for known and fixed $\boldsymbol{\beta}$. Clearly, $\hat{\tau}_k^{x\beta}$ and $\hat{\tau}_{k'}^{x\beta}$ are asymptotically independent because the $K$ subgroups are mutually exclusive (i.e., contain different members). Thus, it follows from these two results that $\hat{\tau}_k^x$ and $\hat{\tau}_{k'}^x$ are also asymptotically independent.

Next, because $\hat{\tau}_k^{x\beta}$ and $\hat{\tau}_{k'}^{x\beta}$ are asymptotically independent and normal, the joint asymptotic distribution of $(\hat{\tau}_1^{x\beta}, \ldots, \hat{\tau}_K^{x\beta})$ is multivariate normal. Thus, it follows by the Cramer-Wold theorem that $(\hat{\tau}_1^x, \ldots, \hat{\tau}_K^x)$ is also jointly asymptotically normal, because for each $(c_1, \ldots, c_K) \in \mathcal{R}^K$, we have that $\sum_{k=1}^{K} c_k \hat{\tau}_k^x \xrightarrow{d} \sum_{k=1}^{K} c_k \hat{\tau}_k^{x\beta}$. Thus, $\left( \frac{\hat{\tau}_1^x - \tau_1}{\sqrt{\mathrm{Var}(\bar{D}_1)}}, \ldots, \frac{\hat{\tau}_K^x - \tau_K}{\sqrt{\mathrm{Var}(\bar{D}_K)}} \right) \xrightarrow{d} N(\mathbf{0}_{K \times 1}, \mathbf{I}_{K \times K})$. ∎

## B. Statistical details for results in Remarks 7 and 9 after Theorem 1

### B.1. Proof of finite sample results in Remark 7

In this section, we prove finite sample results on the unbiasedness and variance of the differences-in-mean estimator, $\hat{\tau}_k$, from (5) in the main text. We follow the approach in Miratrix et al. [4].

**Result 1.** Under (*C1*), (*C2*), and (*C4*), over the randomization distribution $(R)$, the differences-in-mean estimator, $\hat{\tau}_k$, from (5) in the main text is unbiased for $\tau_k$ with variance,

$$\mathrm{Var}_R(\bar{D}_k) = E_A\left(\frac{1}{n_k^1}\right)\Omega_{R_k}^2(1) + E_A\left(\frac{1}{n_k^0}\right)\Omega_{R_k}^2(0) - \frac{\Omega^2(\tau_k)}{n_k}, \tag{B.1}$$

where expectations are taken over all possible subgroup allocations $(A)$ to the treatment and control groups conditional on $n_k^1 > 0$ or $n_k^0 > 0$.

*Proof.* We show that $\hat{\tau}_k$ is unbiased over $R$ by first conditioning on $n_k^1$ (or $n_k^0 = n_k - n_k^1$), and then averaging over possible subgroup allocations $(A)$ of $n_k^1$ values to the treatment and control groups, conditional on $n_k^1 > 0$ or $n_k^0 > 0$ that we do not denote to reduce notation:

$$\begin{aligned}
E_R(\hat{\tau}_k) &= E_A E_R(\hat{\tau}_k | n_k^1) \\
&= E_A\left[\frac{1}{n_k^1}\sum_{i=1}^{n} G_{ik} Y_i(1) E_R(T_i | n_k^1) - \frac{1}{n_k^0}\sum_{i=1}^{n} G_{ik} Y_i(0) E_R((1-T_i) | n_k^1)\right] \\
&= E_A\left[\frac{1}{n_k}\sum_{i=1}^{n} G_{ik} Y_i(1) - \frac{1}{n_k}\sum_{i=1}^{n} G_{ik} Y_i(0)\right] = \tau_k.
\end{aligned} \tag{B.2}$$

The third equality holds because $E_R(T_i | n_k^1) = n_k^1 / n_k$ and $E_R((1-T_i) | n_k^1) = n_k^0 / n_k$. Thus, $\hat{\tau}_k$ is unbiased.

We can similarly calculate the unconditional variance of $\hat{\tau}_k$ using the law of total variance:

$$Var_R(\hat{\tau}_k) = E_A(Var_R(\hat{\tau}_k|n_k^1)) + Var_A(E_R(\hat{\tau}_k|n_k^1)). \tag{B.3}$$

From (B.2), we have that $E_R(\hat{\tau}_k|n_k^1) = \tau_k$, which implies that $Var_A\left(E_R(\hat{\tau}_k|n_k^1)\right) = 0$. Thus, $Var_R(\hat{\tau}_k) = E_A(Var_R(\hat{\tau}_k|n_k^1))$.

To calculate $Var_R(\hat{\tau}_k|n_k^1)$, we apply the full sample design-based results from the literature [5-8], because conditioning on $n_k^1$ yields fixed subgroup treatment and control group sample sizes for the analysis. This approach yields,

$$Var_R(\hat{\tau}_k|n_k^1) = \frac{\Omega_{R_k}^2(1)}{n_k^1} + \frac{\Omega_{R_k}^2(0)}{n_k^0} - \frac{\Omega^2(\tau_k)}{n_k}. \tag{B.4}$$

Finally, if we average (B.4) over $A$ to calculate $E_A(Var_R(\hat{\tau}_k|n_k^1))$, we obtain (B.1). ∎

### B.2. Variance for the weighted nonresponse estimator in Remark 9

To account for data nonresponse using propensity score weights (assumed known), the ATE estimator, $\hat{\tau}_k^{rx}$, using the model (6) in the main text can be obtained using the respondent sample and weighted least squares (WLS). Under the conditions discussed in Remark 9, the resulting variance for the CLT is based on expected subgroup respondent sizes and is as follows:

$$\text{Var}(\bar{D}_k^r) = \phi_k^r \left[\frac{\Omega_{R_k}^{r2}(1)}{n_k p \bar{r}_k} + \frac{\Omega_{R_k}^{r2}(0)}{n_k(1-p)\bar{r}_k} - \frac{\Omega^{r2}(\tau_k)}{n_k \bar{r}_k}\right], \tag{B.5}$$

where $\bar{r}_k = \frac{1}{n_k}\sum_{i=1}^n G_{ik} r_i$ is the response rate for subgroup $k$; $\Omega_{R_k}^{r2}(t) = \frac{1}{(n_k \bar{r}_k - 1)}\sum_{i=1}^n \frac{w_i^2}{(\bar{w}_k^r)^2} \varepsilon_{ik}^{r2}(t)$ are weighted variances; $\bar{w}_k^r = \frac{1}{n_k \bar{r}_k}\sum_{i=1}^n G_{ik} r_i w_i^r$ is the mean respondent weight; $\varepsilon_{ik}^r(t)$ are residuals based on weighted means; $\Omega^{2r}(\tau_k) = \frac{1}{(n_k \bar{r}_k - 1)}\sum_{i=1}^n \frac{w_i^2}{(\bar{w}_k^r)^2}(\varepsilon_{ik}^r(1) - \varepsilon_{ik}^r(0))^2$ is the weighted heterogeneity term; and $\phi_k^r = (n_k \bar{r}_k - 1)/(n_k \bar{r}_k - \pi_k \bar{r}_k)$ is the correction term. The first two denominators in (B.5) are expected subgroup respondent sizes. This variance can be estimated using sample counterparts for the terms in (B.5) and by adapting (12) in the main text.

### C. Comparing design-based and robust variance estimators and additional results for the clustered RCT

In this section, we first compare the design-based variance estimator in (12) for the non-clustered design to the robust Huber-White (HW) variance estimator [9,10]. We then display a consistent design-based variance estimator for the clustered RCT and compare it to the cluster-robust standard error (CRSE) estimator [11]. Finally, we discuss additional subgroup results for the clustered RCT.

## C.1. Comparing the design-based and HW estimators for the non-clustered design

The HW estimator for the OLS coefficients in (6) in the main text is,

$$\text{V\^ar}(\hat{\boldsymbol{\gamma}}_{HW}) = g(\sum_{i=1}^{n} \tilde{\mathbf{z}}_i' \tilde{\mathbf{z}}_i)^{-1}(\sum_{i=1}^{n} \tilde{\mathbf{z}}_i' \tilde{\mathbf{z}}_i \hat{e}_i^2)(\sum_{i=1}^{n} \tilde{\mathbf{z}}_i' \tilde{\mathbf{z}}_i)^{-1}, \tag{C.1}$$

where $\tilde{\mathbf{z}}_i$ is a $1xl$ vector of explanatory variables, $\hat{e}_i$ are OLS residuals, and $g$ is a small sample correction term. Here, we use $g = n/(n-l)$, the default value in Stata and R.

To highlight differences between (12) in the main text and (C.1), consider the model without covariates, where $\tilde{\mathbf{z}}_i = (G_{i1}\tilde{T}_i, \dots, G_{iK}\tilde{T}_i, G_{i1}, \dots, G_{iK})$. Note that $\tilde{\mathbf{z}}_i'\tilde{\mathbf{z}}_i$ is not block diagonal. However, if we apply the asymptotic result, $\frac{1}{n}\sum_{i=1}^{n} G_{ik}\tilde{T}_i = \frac{n^1 n^0}{(n)^2}(\pi_k^1 - \pi_k^0) \xrightarrow{p} 0$ as is done in (12), then $\tilde{\mathbf{z}}_i'\tilde{\mathbf{z}}_i$ becomes block diagonal and the HW ATE estimator for subgroup $k$ in (C.1) is,

$$\text{V\^ar}(\hat{\tau}_{k,HW}) = \left(\frac{n}{n-2K}\right)\frac{(1-p)^2 n_k^1 s_{R_k}^{2*}(1) + p^2 n_k^0 s_{R_k}^{2*}(0)}{[(1-p)^2 n_k^1 + p^2 n_k^0]^2}, \tag{C.2}$$

where $s_{R_k}^{2*}(t) = \frac{(n_k^t - 1)}{n_k^t} s_{R_k}^2(t)$. If we further apply the large sample results, $n_k^1 \approx n_k p$ and $n_k^0 \approx n_k(1-p)$, then (C.2) reduces to,

$$\text{V\^ar}(\hat{\tau}_{k,HW}) = \left(\frac{n}{n-2K}\right)\left[\frac{s_{R_k}^{2*}(1)}{n_k p} + \frac{s_{R_k}^{2*}(0)}{n_k(1-p)}\right], \tag{C.3}$$

compared to the design-based estimator in (12),

$$\text{V\^ar}(\bar{D}_k) = \left(\frac{n_k^1}{n_k^1 - 1}\right)\frac{s_{R_k}^{2*}(1)}{n_k p} + \left(\frac{n_k^0}{n_k^0 - 1}\right)\frac{s_{R_k}^{2*}(0)}{n_k(1-p)}. \tag{C.4}$$

The two estimators are asymptotically equivalent. In finite samples, however, the *df* corrections differ, because the HW estimator applies a global correction based on $n$, whereas the design-based estimator applies a separate correction for each research group using the subgroup sizes. This will typically lead to larger variances for the design-based estimator.

For the model with covariates, the same approach can be used to show that the design-based and HW estimators are asymptotically equivalent. However, it is more complex to make general statements about the relative sizes of the two regression-adjusted estimators in finite samples.

## C.2. Comparing the design-based and CRSE estimators for the clustered design

As discussed in Section 4.2 of the main text, as $m \to \infty$ for the clustered RCT, the full sample design-based CLT results in Schochet et al. [1] can be applied to the ATE estimator for an individual-level subgroup, $\hat{\tau}_{k,clus}^x$, whose members are present in each cluster. This can be done by invoking the regularity conditions in [1] for each subgroup using the cluster-level residuals, $R_{jk}(t) = \frac{n_{jk}}{\bar{n}_k}(\bar{Y}_{jk}(t) - \bar{\bar{Y}}_k(t) - (\bar{x}_{jk} - \bar{\bar{x}}_k)\beta) = \frac{n_{jk}}{\bar{n}_k}\bar{\varepsilon}_{jk}(t)$, where $\bar{n}_k = \frac{1}{m}\sum_{j=1}^{m} n_{jk}$ is the mean subgroup size (mean weight) per cluster.

This approach yields the following CLT result: $\frac{1}{\sqrt{\text{Var}(\bar{\bar{D}}_{k,clus})}}(\hat{\tau}_{k,clus}^x - \tau_k) \xrightarrow{d} N(0,1)$, where $\text{Var}(\bar{\bar{D}}_{k,clus})$ is defined using cluster-level subgroup residuals as follows:

$$\text{Var}(\bar{\bar{D}}_{k,clus}) = \frac{\bar{S}_{R_k}^2(1)}{m^1} + \frac{\bar{S}_{R_k}^2(0)}{m^0} - \frac{\bar{S}^2(\tau_k)}{m}, \tag{C.5}$$

where $\bar{S}_{R_k}^2(t) = \frac{1}{(m-1)}\sum_{j=1}^{m} \frac{n_{jk}^2}{\bar{n}_k^2} \bar{\varepsilon}_{jk}^2(t)$ and $\bar{S}^2(\tau_k) = \frac{1}{(m-1)}\sum_{j=1}^{m} \frac{n_{jk}^2}{\bar{n}_k^2} (\bar{\varepsilon}_{jk}(1) - \bar{\varepsilon}_{jk}(0))^2$.

The variance in (C.5) can be consistently estimated using a version of (12) in the main text based on estimated cluster-level residuals, where the *df* adjustments reflect the number of clusters, not persons:

$$\hat{\text{Var}}(\bar{\bar{D}}_{k,clus}) = \frac{\bar{s}_{R_k}^2(1)}{m^1} + \frac{\bar{s}_{R_k}^2(0)}{m^0}, \tag{C.6}$$

where

$$\bar{s}_{R_k}^2(1) = \frac{1}{(m^1 - Vp\pi_{k,clus}^1 - 1)} \sum_{j=1}^{m} T_j (\bar{y}_{jk} - \hat{\alpha}_{k,clus}^x - (1-p)\hat{\tau}_{k,clus}^x - \tilde{\bar{x}}_{jk}\hat{\beta})^2$$

and

$$\bar{s}_{R_k}^2(0) = \frac{1}{(n_k^0 - V(1-p)\pi_{k,clus}^0 - 1)} \sum_{j=1}^{m} (1-T_j)(\bar{y}_{jk} - \hat{\alpha}_{k,clus}^x + p\hat{\tau}_{k,clus}^x - \tilde{\bar{x}}_{jk}\hat{\beta})^2,$$

where $\bar{y}_{jk} = \frac{1}{n_{jk}}\sum_{i=1}^{n_j} G_{ijk} y_{ij}$ is the mean subgroup outcome, and similarly for $\tilde{\bar{x}}_{jk}$.

The CRSE estimator for the clustered RCT using the OLS coefficients in (6) is,

$$\hat{\text{Var}}(\hat{\gamma}_{CRSE}) = g_{clus}(\tilde{z}'\tilde{z})^{-1} \left(\sum_{j=1}^{m} \tilde{z}_j' \hat{e}_j \hat{e}_j' \tilde{z}_j\right) (\tilde{z}'z)^{-1}, \tag{C.7}$$

where $\tilde{z}$ is an $nxl$ matrix with the full set of explanatory variables; $\tilde{z}_j$ is an $n_j xl$ vector of explanatory variables for those in cluster $j$; $\hat{e}_j$ is a vector of OLS residuals for each cluster, and $g_{clus}$ is a small sample correction. Here, we use $g_{clus} = \left(\frac{m}{m-1}\right)\left(\frac{n-1}{n-l}\right)$, a common value in statistical software packages such as Stata.

To highlight differences between the design-based and CRSE variance estimators, we follow the same approach as for the HW estimator, where we focus on the clustered RCT without covariates. First, we apply the same asymptotic approximations as for the HW estimator to generate a block diagonal $\tilde{z}'\tilde{z}$ matrix in (C.7) and produce the following subgroup ATE estimator using (C.7):

$$\hat{\text{Var}}(\hat{\tau}_{k,CRSE}) = \left(\frac{m}{m-1}\right)\left(\frac{n-1}{n-2K}\right)\frac{(1-p)^2(m^1-1)(\bar{n}_k^1)^2\bar{s}_{R_k}^2(1) + p^2(m^0-1)(\bar{n}_k^0)^2\bar{s}_{R_k}^2(0)}{[(1-p)^2 m^1 \bar{n}_k^1 + p^2 m^0 \bar{n}_k^0]^2}, \quad (C.8)$$

where $\bar{n}_k^t = \frac{1}{m^t}\sum_{j:T_j=t} n_{jk}$ is the mean cluster subgroup size. If we further apply the asymptotic approximations, $\bar{n}_k^1 \approx \bar{n}_k$ and $\bar{n}_k^0 \approx \bar{n}_k$, where $\bar{n}_k = \frac{1}{m}\sum_{j=1}^m n_{jk}$, then (C.8) reduces to,

$$\hat{\text{Var}}(\hat{\tau}_{k,CRSE}) = \left(\frac{m}{m-1}\right)\left(\frac{n-1}{n-2K}\right)\left[\frac{\bar{s}_{R_k}^{2*}(1)}{m^1} + \frac{\bar{s}_{R_k}^{2*}(0)}{m^0}\right], \quad (C.9)$$

where $\bar{s}_{R_k}^{2*}(t) = \frac{(m^t-1)}{m^t}\bar{s}_{R_k}^2(t)$, compared to the design-based estimator in (C.6),

$$\hat{\text{Var}}(\bar{D}_k) = \left(\frac{m^1}{m^1-1}\right)\frac{\bar{s}_{R_k}^{2*}(1)}{m^1} + \left(\frac{m^0}{m^0-1}\right)\frac{\bar{s}_{R_k}^{2*}(0)}{m^0}. \quad (C.10)$$

As with the HW estimator, the two estimators are asymptotically equivalent. In finite samples, however, the *df* corrections differ, which will typically lead to larger variances for the design-based estimator for the model without covariates. The same approach can be used to show that the regression-adjusted estimators are also asymptotically equivalent.

### C.3. Analysis of cluster-level subgroups for the clustered design

Consider a subgroup defined by a cluster-level characteristic ($G_{jk} = 1$), such as a school performance category. In this setting, $m_k^1$ and $m_k^0$ become random, which parallels the subgroup analysis for the non-clustered design. Let $m_k = \sum_{j=1}^m G_{jk}$ denote the number of clusters in subgroup $k$, where $\pi_{k,clus} = m_k/m$ is the population share. We assume that $\pi_{k,clus}$ converges to $\pi_{k,clus}^*$ as $m \to \infty$ for $0 < \pi_{k,clus}^* < 1$ and $\sum_{k=1}^K \pi_{k,clus}^* = 1$. Then, the CLT results for the clustered design in Schochet et al. [1] still apply using the weights, $w_{jk} = G_{jk}n_j$. In this case, the variance for the CLT is,

$$\text{Var}(\bar{\bar{D}}_{k,clus}) = \phi_{k,clus}\left[\frac{\bar{\Omega}_{R_k}^2(1)}{m_k p} + \frac{\bar{\Omega}_{R_k}^2(0)}{m_k(1-p)} - \frac{\bar{\Omega}^2(\tau_k)}{m_k}\right], \quad (C.11)$$

where $\bar{\Omega}_{R_k}^2(t) = \frac{1}{(m_k-1)}\sum_{j=1}^m \frac{G_{jk}n_j^2}{\bar{w}_k^{*2}}\bar{\varepsilon}_{jk}^{*2}(t)$ is the weighted variance; $\bar{w}_k^{*2} = \frac{1}{m_k}\sum_{j=1}^m G_{jk}n_j^2$ is the mean subgroup weight; $\bar{\varepsilon}_{jk}^*(t)$ is the weighted cluster-level residual based on $w_{jk}$; $\bar{\Omega}^2(\tau_k) = \frac{1}{(m_k-1)}\sum_{j=1}^m \frac{G_{jk}n_j^2}{\bar{w}_k^{*2}}(\bar{\varepsilon}_{jk}^*(1) - \bar{\varepsilon}_{jk}^*(0))^2$; and $\phi_{k,clus} = (m_k-1)/(m_k - \pi_{k,clus})$ is the correction.

## D. Additional simulation results

Tables D.1 to D.4 present additional simulation results for the subgroup ATE estimators for various model specifications. Tables D.1 and D.2 provide more detailed findings for the specifications from Table 1 in the main text for the non-clustered RCT, including Type 1 errors for t-tests and F-tests to gauge the difference between the Subgroup 1 and 2 effects (which are both truly zero).

Table D.3 presents selected results for several variants of the variance formula in (12) in the main text. First, we subtract $\frac{1}{n_k}\left(s_{R_k}(1) - s_{R_k}(0)\right)^2$, a lower bound on the heterogeneity (finite-population) term, $\frac{1}{n_k}\Omega^2(\tau_k)$, based on the Cauchy-Schwarz inequality (sharper bounds are discussed in [12]). Second, for models with covariates, we multiply (12) by $(1 - R^2_{TXk})^{-1}$, where $R^2_{TXk}$ is the $R^2$ from a regression of $G_{ik}\tilde{T}_i$ on $\tilde{\mathbf{x}}_{ik}$ and the other terms in (6) to capture finite sample collinearities. Third, for hypothesis testing, we apply the following degrees of freedom from [13], a special case of the Welch 1951 correction [14], that we adapt to our regression-adjusted estimators:

$$df = \frac{(n_k^1 + n_k^0)^2(n_k^1 - Vp\pi_k^1 - 1)(n_k^0 - V(1-p)\pi_k^0 - 1)}{(n_k^1)^2(n_k^1 - Vp\pi_k^1 - 1) + (n_k^0)^2(n_k^0 - V(1-p)\pi_k^0 - 1)}. \tag{D.1}$$

This approach gives more weight to the research group with the smaller subgroup sample size, because it is less precisely estimated than the larger group [15]. Fourth, we apply a $\phi_k$ correction in (12) by subtracting $\pi_k^t$ from the denominator of $s^2_{R_k}(t)$ rather than 1. Finally, we present results using chi-squared distributions to generate the model covariates and errors, with the same means and variances as for the benchmark normal distributions. This approach allows for distribution skewness.

We find that the variants in Table D.3 do not improve simulation performance. Subtracting the lower bound on the heterogeneity term slightly reduces over-coverage of the confidence intervals but multiplying by $(1 - R^2_{TXk})^{-1}$, which performs well for clustered RCTs [1], or using the Bell and McCaffery *df* correction increases over-coverage. However, these differences are small. We find also that as predicted by theory, using the finite-sample variance formula in (11) yields larger standard errors than using the expected sample sizes, but again the differences are small.

Finally, Table D.4 presents selected results for the model with covariates using $p = .4$ and $.6$ rather than the $p = .5$ value used in the other tables. The overall findings remain unchanged for these specifications.

**Table D.1.** Full simulation results for the model without covariates.

| Model specification | Bias[a] | CI coverage | True SE[a,b] | Mean estimated SE | Type 1 error t-test for Subgroup 1 | Type 1 error F-test for difference |
|---|---|---|---|---|---|---|
| **Model without covariates** | | | | | | |
| <u>Sample size: $n = 40, \pi_1 = .50$</u> | | | | | | |
| Design-based (DB), actual subgroup sizes, $\phi_k = 1$ | -.002 | .954 | .646 | .640 | .046 | .047 |
| DB, expected sizes, $\phi_k = 1$ | -.002 | .952 | .646 | .633 | .048 | .050 |
| Huber-White (HW) | -.002 | .953 | .646 | .639 | .047 | .048 |
| <u>Sample size: $n = 40, \pi_1 = .75$</u> | | | | | | |
| DB, actual sizes, $\phi_k = 1$ | .001 | .955 | .532 | .530 | .045 | .059 |
| DB, expected sizes, $\phi_k = 1$ | .001 | .955 | .532 | .528 | .045 | .060 |
| HW | .001 | .959 | .532 | .539 | .041 | .067 |
| <u>Sample size: $n = 100, \pi_1 = .50$</u> | | | | | | |
| DB, actual sizes, $\phi_k = 1$ | .000 | .958 | .376 | .385 | .042 | .042 |
| DB, expected sizes, $\phi_k = 1$ | .000 | .957 | .376 | .383 | .043 | .043 |
| HW | .000 | .958 | .376 | .385 | .042 | .042 |
| <u>Sample size: $n = 100, \pi_1 = .25$</u> | | | | | | |
| DB, actual sizes, $\phi_k = 1$ | .000 | .953 | .626 | .628 | .047 | .049 |
| DB, expected sizes, $\phi_k = 1$ | .000 | .951 | .626 | .618 | .049 | .052 |
| HW | .000 | .948 | .626 | .613 | .052 | .053 |
| <u>Sample size: $n = 100, \pi_1 = .75$</u> | | | | | | |
| DB, actual sizes, $\phi_k = 1$ | .002 | .957 | .325 | .334 | .043 | .048 |
| DB, expected sizes, $\phi_k = 1$ | .002 | .957 | .325 | .333 | .043 | .051 |
| HW | .002 | .959 | .325 | .336 | .041 | .052 |
| <u>Sample size: $n = 200, \pi_1 = .50$</u> | | | | | | |
| DB, actual sizes, $\phi_k = 1$ | .000 | .956 | .289 | .287 | .044 | .043 |
| DB, expected sizes, $\phi_k = 1$ | .000 | .956 | .289 | .286 | .044 | .044 |
| HW | .000 | .956 | .289 | .287 | .044 | .043 |
| <u>Sample size: $n = 200, \pi_1 = .25$</u> | | | | | | |
| DB, actual sizes, $\phi_k = 1$ | .004 | .956 | .421 | .407 | .045 | .046 |
| DB, expected sizes, $\phi_k = 1$ | .004 | .954 | .421 | .404 | .046 | .047 |
| HW | .004 | .954 | .421 | .403 | .047 | .048 |
| <u>Sample size: $n = 200, \pi_1 = .75$</u> | | | | | | |
| DB, actual sizes, $\phi_k = 1$ | .000 | .956 | .235 | .233 | .044 | .045 |
| DB, expected sizes, $\phi_k = 1$ | .000 | .956 | .235 | .233 | .044 | .046 |
| HW | .000 | .957 | .235 | .234 | .043 | .047 |

Notes. See text for simulation details. The calculations assume two subgroups with a focus on results for Subgroup 1, a treatment assignment rate of $p = .50$, and normally distributed covariates and errors. For each specification, the figures are based on 10,000 simulations for each of 5 potential outcome draws, and the findings average across the 5 draws. Ordinary least square (OLS) methods are used for ATE estimation using the model in (6) in the main text, and design-based standard errors are obtained using (12). Huber-White estimates are obtained using the lm_robust procedure in R.

ATE = Average treatment effect; CI = Confidence interval; SE = Standard error; DB = Design-based; HW = Huber-White.

[a] Biases and true standard errors are the same for all specifications within each sample size grouping as they use the same data and OLS model for ATE estimation.

[b] True standard errors are measured as the standard deviation of the estimated treatment effects across simulations.

**Table D.2.** Full simulation results for the model with two covariates.

| Model specification | Bias[a] | CI coverage | True SE[a,b] | Mean estimated SE | Type 1 error t-test for Subgroup 1 | Type 1 error F-test for difference |
|---|---|---|---|---|---|---|
| **Model with two covariates** | | | | | | |
| Sample size: $n = 40, \pi_1 = .50$ | | | | | | |
| DB, actual sizes, $\phi_k = 1$ | .002 | .950 | .501 | .482 | .050 | .050 |
| DB, expected sizes, $\phi_k = 1$ | .002 | .948 | .501 | .476 | .052 | .052 |
| HW | .002 | .953 | .501 | .494 | .047 | .047 |
| Sample size: $n = 40, \pi_1 = .75$ | | | | | | |
| DB, actual sizes, $\phi_k = 1$ | .000 | .953 | .401 | .383 | .047 | .057 |
| DB, expected sizes, $\phi_k = 1$ | .000 | .952 | .401 | .381 | .048 | .059 |
| HW | .000 | .963 | .401 | .406 | .037 | .066 |
| Sample size: $n = 100, \pi_1 = .50$ | | | | | | |
| DB, actual sizes, $\phi_k = 1$ | .000 | .961 | .298 | .305 | .039 | .043 |
| DB, expected sizes, $\phi_k = 1$ | .000 | .960 | .298 | .303 | .040 | .044 |
| HW | .000 | .962 | .298 | .308 | .038 | .041 |
| Sample size: $n = 100, \pi_1 = .25$ | | | | | | |
| DB, actual sizes, $\phi_k = 1$ | .000 | .956 | .486 | .482 | .044 | .047 |
| DB, expected sizes, $\phi_k = 1$ | .000 | .954 | .486 | .475 | .046 | .049 |
| HW | .000 | .952 | .486 | .470 | .048 | .050 |
| Sample size: $n = 100, \pi_1 = .75$ | | | | | | |
| DB, actual sizes, $\phi_k = 1$ | .002 | .959 | .233 | .242 | .041 | .048 |
| DB, expected sizes, $\phi_k = 1$ | .002 | .959 | .233 | .242 | .041 | .050 |
| HW | .002 | .963 | .233 | .247 | .037 | .052 |
| Sample size: $n = 200, \pi_1 = .50$ | | | | | | |
| DB, actual sizes, $\phi_k = 1$ | .000 | .958 | .236 | .232 | .042 | .040 |
| DB, expected sizes, $\phi_k = 1$ | .000 | .957 | .236 | .232 | .043 | .040 |
| HW | .000 | .959 | .236 | .233 | .041 | .039 |
| Sample size: $n = 200, \pi_1 = .25$ | | | | | | |
| DB, actual sizes, $\phi_k = 1$ | .004 | .960 | .347 | .328 | .041 | .045 |
| DB, expected sizes, $\phi_k = 1$ | .004 | .958 | .347 | .325 | .042 | .046 |
| HW | .004 | .958 | .347 | .324 | .043 | .046 |
| Sample size: $n = 200, \pi_1 = .75$ | | | | | | |
| DB, actual sizes, $\phi_k = 1$ | .000 | .960 | .183 | .180 | .040 | .042 |
| DB, expected sizes, $\phi_k = 1$ | .000 | .960 | .183 | .180 | .040 | .043 |
| HW | .000 | .962 | .183 | .182 | .039 | .043 |

Notes. See text for simulation details. The calculations assume two subgroups with a focus on results for Subgroup 1, a treatment assignment rate of $p = .50$, and normally distributed covariates and errors. For each specification, the figures are based on 10,000 simulations for each of 5 potential outcome draws, and the findings average across the 5 draws. Ordinary least square (OLS) methods are used for ATE estimation using the model in (6) in the main text, and design-based standard errors are obtained using (12). Huber-White estimates are obtained using the lm_robust procedure in R.

ATE = Average treatment effect; CI = Confidence interval; SE = Standard error; DB = Design-based; HW = Huber-White.

[a] Biases and true standard errors are the same for all specifications within each sample size grouping because they use the same data and OLS model for ATE estimation.

[b] True standard errors are measured as the standard deviation of the estimated treatment effects across simulations.

**Table D.3.** Simulation results for alternative model specifications.

| Model specification | Confidence interval coverage | True standard error[a] | Mean estimated standard error |
|---|---|---|---|
| **Model without covariates** | | | |
| Sample size: $n = 100, \pi_1 = .5$ | | | |
| Design-based (DB), actual subgroup sizes | .959 | .376 | .385 |
| Finite-population (FP) heterogeneity correction | .957 | .376 | .382 |
| Bell and McCaffery (BM) *df* correction | .958 | .376 | .385 |
| Chi-squared distribution for errors and covariates | .957 | .404 | .403 |
| Finite-sample (FS) variance in (11) | .958 | .376 | .385 |
| Sample size: $n = 100, \pi_1 = .25$ | | | |
| DB, actual sizes | .953 | .626 | .628 |
| FP heterogeneity correction | .950 | .626 | .619 |
| BM *df* correction | .955 | .626 | .628 |
| Chi-squared distribution | .955 | .427 | .432 |
| FS variance in (11) | .954 | .626 | .628 |
| **Model with two covariates** | | | |
| Sample size: $n = 100, \pi_1 = .5$ | | | |
| DB, actual sizes | .961 | .298 | .305 |
| FP heterogeneity correction | .959 | .298 | .301 |
| $R^2_{TXk}$ correction | .964 | .298 | .310 |
| BM *df* correction | .961 | .298 | .305 |
| Chi-squared distribution | .959 | .314 | .308 |
| FS variance in (11) | .961 | .298 | .305 |
| Sample size: $n = 100, \pi_1 = .25$ | | | |
| DB, actual sizes | .956 | .486 | .482 |
| FP heterogeneity correction | .953 | .486 | .474 |
| $R^2_{TXk}$ correction | .962 | .486 | .496 |
| BM *df* correction | .958 | .486 | .482 |
| Chi-squared distribution | .956 | .362 | .366 |
| FS variance in (11) | .958 | .486 | .483 |

Notes. See text for simulation details. The calculations assume two subgroups with a focus on results for Subgroup 1, a treatment assignment rate of $p = .50$, and normally distributed covariates and errors, except for the chi-squared entries. For each specification, the figures are based on 10,000 simulations for each of 5 potential outcome draws, and the findings average across the 5 draws. Ordinary least square (OLS) methods are used for ATE estimation using the model in (6) in the main text, and design-based standard errors are obtained using (12) using actual sample sizes except for the expected size and finite-sample variance entries.

ATE = Average treatment effect; DB = Design-based; FS = Finite-sample.

[a] True standard errors are the same for all specifications within each sample size category except for the chi-squared entries because they use the same data and OLS model for ATE estimation. They are measured as the standard deviation of the estimated treatment effects across simulations.

**Table D.4.** Simulation results for the subgroup ATE estimators for $p = .4$ and $.6$.

| Model specification | Bias of ATE estimator[a] | Confidence interval coverage | True standard error[a,b] | Mean estimated standard error |
|---|---|---|---|---|
| **Model without covariates: $p = .4$** | | | | |
| Sample size: $n = 40, \pi_1 = .5$ | | | | |
| Design-based (DB), actual subgroup sample sizes | .001 | .950 | .641 | .629 |
| DB, expected sizes | .001 | .947 | .641 | .620 |
| Huber-White (HW) | .001 | .948 | .641 | .623 |
| Sample size: $n = 100, \pi_1 = .5$ | | | | |
| DB, actual sizes | .001 | .954 | .403 | .404 |
| DB, expected sizes | .001 | .953 | .403 | .402 |
| HW | .001 | .954 | .403 | .403 |
| Sample size: $n = 100, \pi_1 = .25$ | | | | |
| DB, actual sizes | -.002 | .952 | .611 | .593 |
| DB, expected sizes | -.002 | .949 | .611 | .583 |
| HW | -.002 | .946 | .611 | .576 |
| **Model without covariates: $p = .6$** | | | | |
| Sample size: $n = 40, \pi_1 = .5$ | | | | |
| DB, actual sizes | .000 | .955 | .597 | .588 |
| DB, expected sizes | .000 | .953 | .597 | .581 |
| HW | .000 | .953 | .597 | .585 |
| Sample size: $n = 100, \pi_1 = .5$ | | | | |
| DB, actual sizes | .000 | .956 | .422 | .415 |
| DB, expected sizes | .000 | .955 | .422 | .412 |
| HW | .000 | .956 | .422 | .414 |
| Sample size: $n = 100, \pi_1 = .25$ | | | | |
| DB, actual sizes | .002 | .955 | .530 | .525 |
| DB, expected sizes | .002 | .952 | .530 | .516 |
| HW | .002 | .950 | .530 | .511 |

Notes. See text for simulation details. The calculations assume two subgroups with a focus on results for Subgroup 1, a treatment assignment rate of $p = .4$ or $p = .6$, and normally distributed covariates and errors. For each specification, the figures are based on 10,000 simulations for each of 5 potential outcome draws, and the findings average across the 5 draws. Ordinary least square (OLS) methods are used for ATE estimation using the model in (6) in the main text, and design-based standard errors are obtained using (12). Huber-White estimates are obtained using the lm_robust procedure in R.

ATE = Average treatment effect; DB = Design-based; HW = Huber-White.

[a] Biases and true standard errors are the same for all specifications within each sample size category because they use the same data and OLS model for ATE estimation.

[b] True standard errors are measured as the standard deviation of the estimated treatment effects across simulations.